\documentclass[letter]{aa} 

\usepackage{graphicx}
\usepackage{txfonts}
\begin{document}

   \title{The VISCACHA survey}

   \subtitle{III. Star clusters counterpart of the Magellanic Bridge and Counter-Bridge in 8D}

   \author{B. Dias 
          \inst{1}
          \and
          M. S. Angelo\inst{2} 
          \and
          R. A. P. Oliveira\inst{3} 
          \and
          F. Maia\inst{4} 
          \and
          M. C. Parisi\inst{5,6} 
          \and
          B. De Bortoli\inst{7,8} 
          \and 
          S. O. Souza\inst{3} 
          \and
          O. J. Katime Santrich\inst{9} 
          \and
          L. P. Bassino\inst{7,8} 
          \and
          B. Barbuy\inst{3} 
          \and
          E. Bica\inst{10} 
          \and 
          D. Geisler\inst{11,12,13} 
          \and
          L. Kerber\inst{9} 
          \and
          A. P\'erez-Villegas\inst{3,14}  
          \and
          B. Quint\inst{15} 
          \and
          D. Sanmartim\inst{16} 
          \and
          J. F. C. Santos Jr.\inst{17} 
          \and
          P. Westera\inst{18} 
          }

   \institute{
   Instituto de Alta Investigaci\'on, Sede Esmeralda, Universidad de Tarapac\'a, Av. Luis Emilio Recabarren 2477, Iquique, Chile\\
              \email{bdiasm@academicos.uta.cl}
         \and
    Centro Federal de Educa\c c\~ao Tecnol\'ogica de Minas Gerais, Av. Monsenhor Luiz de Gonzaga, 103, 37250-000 Nepomuceno, MG, Brazil
        \and
    Universidade de S\~ao Paulo, IAG, Rua do Mat\~ao 1226, Cidade Universit\'aria, S\~ao Paulo 05508-900, Brazil
        \and
    Instituto de F\'isica, Universidade Federal do Rio de Janeiro, 21941-972, Rio de Janeiro, RJ, Brazil
    \and
    Observatorio Astron\'omico, Universidad Nacional de C\'ordoba, Laprida 854, X5000BGR, C\'ordoba, Argentina.
    \and
    Instituto de Astronom{\'\i}a Te\'orica y Experimental (CONICET-UNC), Laprida 854, X5000BGR, C\'ordoba, Argentina.
    \and
    Facultad de Ciencias Astronómicas y Geofísicas de la Universidad Nacional de La Plata, and Instituto de Astrofísica de La Plata (CCT La Plata - CONICET, UNLP), Paseo del Bosque S/N, B1900FWA La Plata, Argentina
    \and
    Consejo Nacional de Investigaciones Científicas y Técnicas, Godoy Cruz 2290, C1425FQB, Ciudad Autónoma de Buenos Aires, Argentina
    \and 
    Departamento de Ci\^encias Exatas e Tecnol\'ogicas, UESC, Rodovia Jorge Amado km 16, 45662-900, Brazil 
    \and
    Departamento de Astronomia, IF - UFRGS, Av. Bento Gon\c calves 9500, 91501-970, Brazil 
    \and
    Departamento de Física y Astronomía, Universidad de La Serena, Avenida Juan Cisternas 1200, La Serena, Chile
    \and
    Instituto de Investigación Multidisciplinario en Ciencia y Tecnología, Universidad de La Serena Benavente 980, La Serena, Chile
    \and
    Departmento de Astronomía, Universidad de Concepción, Casilla 160-C, Concepción, Chile
    \and
    Instituto de Astronomía, Universidad Nacional Autonóma de México, Apartado Postal 106, C.P. 22800, Ensenada, Baja California, Mexico
    \and
    Gemini Observatory/NSF’s NOIRLab, Casilla 603, La Serena, Chile
    \and
    Carnegie Observatories, Las Campanas Observatory, Casilla 601, La Serena, Chile
    \and
    Departamento de F\'isica, ICEx - UFMG, Av. Ant\^onio Carlos 6627, Belo Horizonte 31270-901, Brazil
    \and
    Universidade Federal do ABC, Centro de Ci\^encias Naturais e Humanas, Avenida dos Estados, 5001, 09210-580, Brazil
    }

   \date{Received ; accepted }

 
  \abstract
   {The interactions between the Small and Large Magellanic Clouds (SMC and LMC) created the Magellanic Bridge; a stream of gas and stars pulled out of the SMC towards the LMC about 150\,Myr ago. The tidal counterpart of this structure, which should include a trailing arm, has been predicted by models but no compelling observational evidence has confirmed the Counter-Bridge so far.}
   {The main goal of this work is to find the stellar counterpart of the Magellanic Bridge and Counter-Bridge. We use star clusters in the SMC outskirts as they provide a 6D phase-space vector, age, and metallicity which help characterise the outskirts of the SMC.
   }
   {Distances, ages, and photometric metallicities were derived from fitting isochrones to the colour-magnitude diagrams from the VISCACHA survey. Radial velocities and spectroscopic metallicities were derived from the spectroscopic follow-up using GMOS in the CaII triplet region.}
   {Among the seven clusters analysed in this work, five belong to the Magellanic Bridge, one belongs to the Counter-Bridge, and the other belongs to the transition region.}
   {The existence of the tidal counterpart of the Magellanic Bridge is evidenced by star clusters. The stellar component of the Magellanic Bridge and Counter-Bridge are confirmed in the SMC outskirts. These results are an important constraint for models that seek to reconstruct the history of the orbit and interactions between the LMC and SMC as well as constrain their future interaction including with the Milky Way.}


   \maketitle

\section{Introduction}
The past interaction history between the Small and Large Magellanic Clouds (SMC, LMC) and of both galaxies with the Milky Way has been a controversial topic of discussion in the last few decades. A canonical scenario describes the SMC and LMC as bound to the Milky Way for more than 5~Gyr, and it prescribes that they only became a pair recently, meaning that the Magellanic Stream and Leading Arm were formed from the interaction among the three galaxies \citep[e.g.][]{gardiner+96,diaz+11,diaz+12}. An alternative scenario describes the Magellanic System as a bound SMC-LMC pair that is in its first infall towards the Milky Way or at least in an elongated orbit with a 6~Gyr period \citep[e.g.][]{besla+07,besla+10}. In both cases, simulations are able to roughly reproduce the large-scale gas structure of the Magellanic Stream and Leading Arm, traced by HI gas \citep[e.g.][]{putman+98,nidever+08}, the Magellanic gaseous bridge \citep[e.g.][]{hindman+63,harris07}, and the old RR Lyrae bridge \citep[e.g.][]{jacyszyn+17,belokurov+17}. However, the smaller structures still need further simulations to fine-tune the most recent observational constraints. This task is particularly challenging for the complex structure of the SMC that is falling apart with a large line-of-sight depth which is not trivial to characterise \citep[e.g.][]{bekki+09,besla+07,beslaPhD,dias+16,niederhofer+18,zivick+18,deleo+20}. More multi-dimensional observational constraints are required to help trace the disruption scenario of the SMC.

An important observed feature of the SMC structure that has been detected by different surveys is the large line-of-sight depth of the stellar distribution in the north-eastern SMC, revealing a population at the mean distance of the SMC at about 62~kpc \citep{degrijs+15,graczyk+20} and a foreground population at about 50~kpc \citep[e.g.][]{nidever+13,mackey+18,omkumar+20}. The foreground component is interpreted as being the stellar counterpart of the Magellanic Bridge (a.k.a. Bridge), detected only between 2.5$^{\circ}$ and 5-6$^{\circ}$ from the SMC towards the LMC. If this is a tidal structure, it would be expected that distance correlates with velocity, which is an indication of what was found by the spectroscopic studies of \cite{dobbie+14} and \cite{deleo+20}. In addition, \cite{pieres+17} also found a stellar over-density towards the north of the SMC, which is also interpreted as a tidally stripped structure.

Various simulations are able to reproduce some of the observed features above, but not entirely; they also make predictions that should be further checked with future observations. For example, \cite{diaz+12} performed $\mathcal{N}$-body simulations, modelling the SMC as a rotating disc (representing the HI gas) and a non-rotating spheroid (representing the old stellar component). They were able to reproduce the Magellanic Stream and Leading Arm, which formed $\sim$2Gyr ago; they also reproduced the Magellanic Bridge (including the bifurcation) and they predicted a tidal counterpart of the Bridge, called the Counter-Bridge, both formed $\sim$150Myr ago (also predicted by other simulations such as those by \citealp{gardiner+96} and \citealp{beslaPhD}, although with a different orientation, shape, extension, and density). \cite{diaz+12} tested three truncation radii to model the spheroid component, the extended model being preferred as it reproduces the morphology and velocity pattern of the Magellanic Bridge better as well as a density break point detected by \cite{nidever+11} at about $7^{\circ}$ from the SMC centre. This is a remnant of a merger event, although with a simulated position at $\sim 4^{\circ}$. On the other hand, this would imply that the stellar counterpart of the Magellanic Bridge is old, which does not seem to be the case of the exclusively young stellar population detected by \cite{harris07}, who concluded that the Magellanic Bridge was formed only by gas, whereas stars were formed in situ. In this case, the compact spheroid should be a better choice, which poses a problem for the simulations. Nevertheless, \cite{harris07} also found intermediate-age stars in the Bridge, but only in regions closer to the LMC and SMC, and they concluded that these stellar components are bound to their respective galaxies and not tidally stripped, because the old stars are confined in the exponential surface density profile of the LMC, before any break point. They did not analyse, in detail, the SMC Bridge population in the same way, presumably because of the complex geometry of the SMC. \cite{nidever+11} applied a similar strategy and found a break point in the density profile coincident with the same threshold of the 'pure-bridge' defined by \cite{harris07} at about RA $\sim 2.5^{h}$. Therefore, the stellar counterpart of the Magellanic Bridge (and Counter-Bridge, as a consequence) is indeed a tidally stripped population from the SMC and may be of an old or intermediate age as proposed by the simulations of \cite{diaz+12}.

The old stellar tracers above are usually red clump stars which are useful for overall maps. A more comprehensive picture can be obtained using star clusters as probes, represented by a 6D phase-space vector, age, and metallicity, building up an 8D map of the SMC outskirts. \cite{glatt+10} derived ages from the colour-magnitude diagrams (CMDs) of 324 SMC clusters, but using shallow photometry from the
MCPS survey \citep[][]{zaritsky+02}, covering only clusters younger than ${\sim} 1$Gyr, and assuming a fixed distance and metallicity. \cite{piatti+15} analysed 51 Bridge clusters from the 
VMC survey \citep[][]{cioni+11} and found a predominant young population
($\sim20$ Myr),
which presumably formed in situ, and an older counterpart 
($\sim1.3$ Gyr),
and they conclude that the Bridge could be older than previously thought. However, they also assumed a fixed distance for all clusters. It remains an open question whether the old population actually corresponds to the foreground tidal stellar counterpart of the Magellanic Bridge. \cite{crowl+01} derived the 3D positions, ages, and metallicities of 12 SMC clusters and found that the eastern SMC side contains younger and more metal-rich clusters and has a larger line-of-sigh depth than the western side, which is in agreement with other works. They endorsed the use of populous star clusters as probes of the SMC structure. 

In this paper we seek to confirm the tidal stripping nature of the SMC Bridge and Counter-Bridge using low-mass star clusters. These clusters have been observed within the VIsible Soar photometry of star Clusters in tApii and Coxi HuguA (VISCACHA) survey\footnote{\url{http://www.astro.iag.usp.br/~viscacha/}} (\citealp[][hereafter Paper I]{maia+19}; \citealp{dias+20}) that has been collecting deep and resolved photometric data for star clusters in the outskirts of the Magellanic Clouds, which are more susceptible to be tidally stripped \citep[e.g.][]{mihos+94}.

\section{Photometric and spectroscopic data}

The VISCACHA photometric data used here were observed in 2016, 2017, and 2019 under programmes SO2016B-018, SO2017B-014, and SO2019B-019, using the ground-based 4m telescope SOAR and its adaptive optics module SAM \citep{tokovinin+16}. Data reduction, PSF photometry, and 
CMD decontamination based on membership probability
were performed following the methodology presented in Paper I.

The spectroscopic data were observed in 2018 with GMOS, Gemini-S, under the Chile-Brazil-Argentina joint programmes GS-2018B-Q-208 and GS-2018B-Q-302 to provide radial velocities (RVs), a crucial parameter to characterise tidal tails. The data reduction was done using default tasks of the Gemini data reduction software, automated by a script developed and fine-tuned by M. S. Angelo\footnote{\url{http://drforum.gemini.edu/topic/gmos-mos-guidelines-part-1/}}. The 1D spectra were extracted, with special attention to the wavelength calibration, that is to say skylines were used as references along with the arc-lamps to enable absolute RV analysis. Proper motions were taken from Gaia Early Data Release 3  \citep[EDR3,][see Appendix \ref{app:vpd}]{gaiaedr3}. 

\subsection{Statistical isochrone fitting}

The CMD morphology of a star cluster depends on the age, metallicity, distance, and reddening. Therefore, it is mandatory to fit all four parameters together to carry out a self-consistent analysis. We used the SIRIUS code \citep{souza+20} that performs a statistical isochrone fitting, with a Bayesian approach based on the Markov chain Monte Carlo sampling method. We adopted the geometrical likelihood function and the membership probability of stars as a uniform prior. The fits are blind to spectroscopic metallicities, which were used to compare with the photometric metallicities as a quality check of both independent analyses.

\subsection{CaT analysis}

Radial velocities were derived by cross-correlation of the observed spectra of individual stars in the clusters against a template synthetic spectrum from the library of \cite{coelho+14} with typical atmospheric parameters for red giant branch (RGB) stars, that is (T$_{\rm eff}$, log($g$), [Fe/H], [$\alpha$/Fe]) = (5000~K, 1.0~dex, -1.0~dex, 0.4~dex). RVs and uncertainties were derived using {\it fxcor} at IRAF, and the conversion from the observed velocities to RV$_{\rm hel}$ was calculated using {\it rvcorrect} at IRAF (Table \ref{tab:results}).
Spectroscopic metallicities were derived following the recipes of \cite{dias+20b}, using a calibration of the reduced equivalent width of CaII triplet lines (CaT) with metallicity.

The membership selection follows the methodology by \cite{parisi+09}. Briefly, stars within the cluster radius with an RV and [Fe/H] within windows of $10{\rm km\cdot s^{-1}}$ and $0.2$dex 
around the group of innermost stars with a common RV and [Fe/H] are considered as members. The cluster radii were taken from \cite{santos+20}, whenever available, or from \cite{bica+20} otherwise.

\begin{table*}[!htb]
    \caption{Derived parameters for the star clusters. Age, [Fe/H]$_{\rm CMD}$, E(B-V), distance from VISCACHA CMD isochrone fitting, [Fe/H]$_{\rm CaT}$, $RV_{hel}$ from GMOS spectra, $\mu_{\alpha}$, and $\mu_{\delta}$ from Gaia EDR3. Distance $a$ follows the definition by \cite{dias+14}, and $ (\alpha,\delta) $ coordinates are from \cite{bica+20}.}
    \label{tab:results}
    \centering
    \scriptsize
    \begin{tabular}{p{1cm}p{0.8cm}p{0.8cm}p{1.8cm}p{0.8cm}p{0.7cm}p{0.5cm}p{1.2cm}p{1.2cm}p{1.5cm}p{1.7cm}p{1.8cm}}
    \hline
    \noalign{\smallskip}
    Cluster & Age & [Fe/H]$_{\rm CMD}$ & [Fe/H]$_{\rm CaT}$ & E(B-V) & Dist. & a & $\alpha_{J2000}$ & $\delta_{J2000}$ &  $RV_{hel}$ & $\mu_{\alpha}\cdot {\rm cos}(\delta)$ & $\mu_{\delta}$ \\
     &  &  & $\pm unc.$(std.dev.) &  &  &  &  &  & $\pm unc.$(std.dev.) & $\pm unc.$(std.dev.) & $\pm unc.$(std.dev.) \\
    name & (Gyr) & (dex) & (dex) & (mag) & (kpc) & (deg) & hh:mm:ss & dd:mm:ss & (${\rm km{\cdot}s^{-1}}$) & (${\rm mas{\cdot}yr^{-1}}$) & (${\rm mas{\cdot}yr^{-1}}$) \\
    \noalign{\smallskip}
    \hline
    \noalign{\smallskip}
    \multicolumn{12}{c}{Magellanic Bridge} \\
    \noalign{\smallskip}
    \hline
    \noalign{\smallskip}
    BS~196 & $3.89^{+0.68}_{-0.50}$ & $-0.75^{+0.22}_{-0.19}$ & $-0.89\pm0.04(0.08)$ & $0.05^{+0.04}_{-0.04}$ & $50.1^{+1.6}_{-2.2}$ & 5.978 & 01:48:01.8 & $-$70:00:13 & $135.5\pm1.4(2.7)$  & $1.12\pm0.07(0.18)$ & $-1.14\pm0.06(0.05)$ \\
    \noalign{\smallskip}
    BS~188* & $1.82^{+0.22}_{-0.20}$ & $-0.58^{+0.13}_{-0.13}$ & $-0.94\pm0.06(0.13)$ & $0.00^{+0.03}_{-0.00}$ & $52.7^{+3.0}_{-3.1}$  & 4.441 & 01:35:10.9 & $-$71:44:11 & $120.3\pm3.5(7.9)$ & $1.25\pm0.08(0.23)$ & $-1.35\pm0.07(0.04)$ \\
    \noalign{\smallskip}
    HW~56** & $3.09^{+0.22}_{-0.14}$ & $-0.54^{+0.07}_{-0.12}$ & $-0.97\pm0.12(0.20)$ & $0.03^{+0.02}_{-0.02}$ & $53.5^{+1.2}_{-1.2}$  & 2.397 & 01:07:41.1    & $-$70:56:06 & $157.7\pm5.4(9.3)$ & $0.99\pm0.11(0.04)$ & $-1.27\pm0.10(0.29)$  \\
    \noalign{\smallskip}
    HW~85 & $1.74^{+0.08}_{-0.12}$ & $-0.83^{+0.07}_{-0.05}$ & $-0.82\pm0.06(0.14)$ & $0.04^{+0.02}_{-0.02}$ & $54.0^{+1.2}_{-2.0}$ & 5.219 & 01:42:28.0 & $-$71:16:44  & $143.2\pm3.0(6.8)$ & $1.26\pm0.09(0.47)$ & $-1.41\pm0.09(0.26)$  \\
    \noalign{\smallskip}
    Lindsay~100 & $3.16^{+0.15}_{-0.14}$ & $-0.73^{+0.03}_{-0.03}$ & $-0.89\pm0.06(0.14)$ & $0.01^{+0.01}_{-0.01}$ & $58.6^{+0.8}_{-0.5}$ & 2.556 & 01:18:16.9 & $-$72:00:06  & $145.8\pm1.4(3.3)$  & $0.98\pm0.05(0.15)$ & $-1.10\pm0.05(0.11)$  \\
    \noalign{\smallskip}
    \hline
    \noalign{\smallskip}
    \multicolumn{12}{c}{Transition} \\
    \noalign{\smallskip}
    \hline
    \noalign{\smallskip}
    Bruck~168  & $6.6^{+0.8}_{-0.9}$ & $-1.22^{+0.20}_{-0.15}$ & $-1.08\pm0.06(0.09)$ & $0.00^{+0.02}_{-0.01}$ & $61.9^{+2.3}_{-2.0}$  & 3.584 & 01:26:42.7    & $-$70:47:01 & $141.7\pm4.6(7.9)$ &  $0.94\pm0.09(0.08)$ & $-1.15\pm0.09(0.04)$ \\
    \noalign{\smallskip}
    \hline
    \noalign{\smallskip}
    \multicolumn{12}{c}{Counter-Bridge} \\
    \noalign{\smallskip}
    \hline
    \noalign{\smallskip}
    IC~1708 & $0.93^{+0.16}_{-0.04}$ & $-1.02^{+0.05}_{-0.10}$ & $-1.11\pm0.06(0.17)$ & $0.06^{+0.02}_{-0.02}$ & $65.2^{+1.2}_{-1.8}$ & 3.286 & 01:24:55.9 & $-$71:11:04 & $214.9\pm2.7(6.6)$  & $0.39\pm0.10(0.35)$ & $-1.26\pm0.07(0.20)$ \\
    \noalign{\smallskip}
    \hline
    \end{tabular}
    \tablefoot{*BS\,188 was observed under sub-optimal weather conditions, and the resulting isochrone fitting is very sensitive to a handful of stars from the shallow CMD, which could explain the metallicity difference between photometric and spectroscopic [Fe/H]. (see Appendix \ref{app:CMDspec}).
    **HW\,56 is immersed in a dense field (a=2.397$^{\circ}$, the smallest in our sample) that has a very similar CMD as the cluster region; therefore, the statistical decontamination left low-probable member stars on the fainter main sequence region, which may bias the photometric metallicities (see Appendix \ref{app:CMDspec}). These uncertainties do not change the conclusions of this work.}
\end{table*}

\section{Discussion}

\begin{figure}[!htb]
    \centering
    \includegraphics[width=0.7\columnwidth]{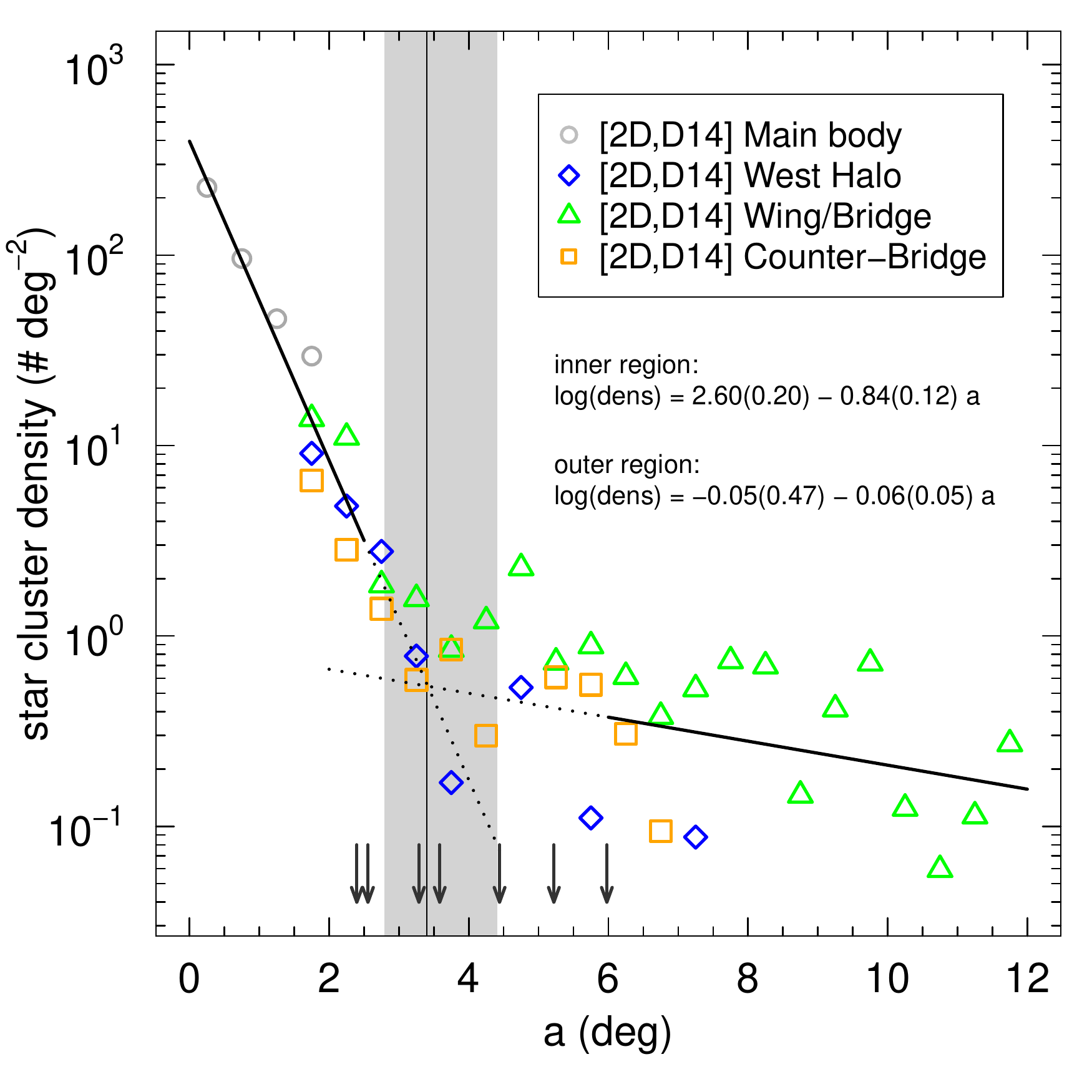}
    \caption{SMC star cluster radial density profile \citep[catalogued by][]{bica+20}, using the semi-major axis $a$ of ellipses drawn around the SMC centre as defined by D14 (see also Fig. \ref{fig:3Dposition}), with $0.5^{\circ}$ bins. Solid lines are the fits to each region, whereas the dotted lines are extrapolation to find the intersection at $a=3.4_{-0.6}^{+1.0}$, represented by the shaded grey $1\sigma$ area. The arrows indicate the position of the seven clusters.}
    \label{fig:SMCprofile}
\end{figure}

The starting point to discuss the SMC tidal tails is to argue whether a particular group of star clusters is bound or unbound to the SMC. This is not trivial; however, \cite{nidever+11} and \cite{diaz+12} have argued that a sharp break in the radial SMC stellar density distribution indicates tidal distortions. 
\citet[][hereafter D14,D16]{dias+14,dias+16} divided the projected distribution of the SMC star cluster population that is outside the SMC main body ($a>2^{\circ}$) into three regions (split by voids in the azimuth direction) possibly related to different disrupting episodes: Wing and Bridge, Counter-Bridge, and West Halo. 
We show in Fig. \ref{fig:SMCprofile} that all regions share a similar radial density profile, with a clear break point at $a=3.4^{\circ}{}_{-0.6}^{+1.0}$, which is consistent with (i) the SMC tidal radius $r_t^{SMC} \sim 4.5^{\circ}$ and the break point for young and old populations found by \cite{massana+20}, (ii) the break point by the simulations of \cite{diaz+12}, and (iii) with the vertex of the V inversion in the metallicity gradient \citep{parisi+15,bica+20}.
All seven clusters analysed in this paper are located around the break point or beyond it
(see Table \ref{tab:results}).

\begin{figure}[!htb]
    \centering
    \includegraphics[width=0.9\columnwidth]{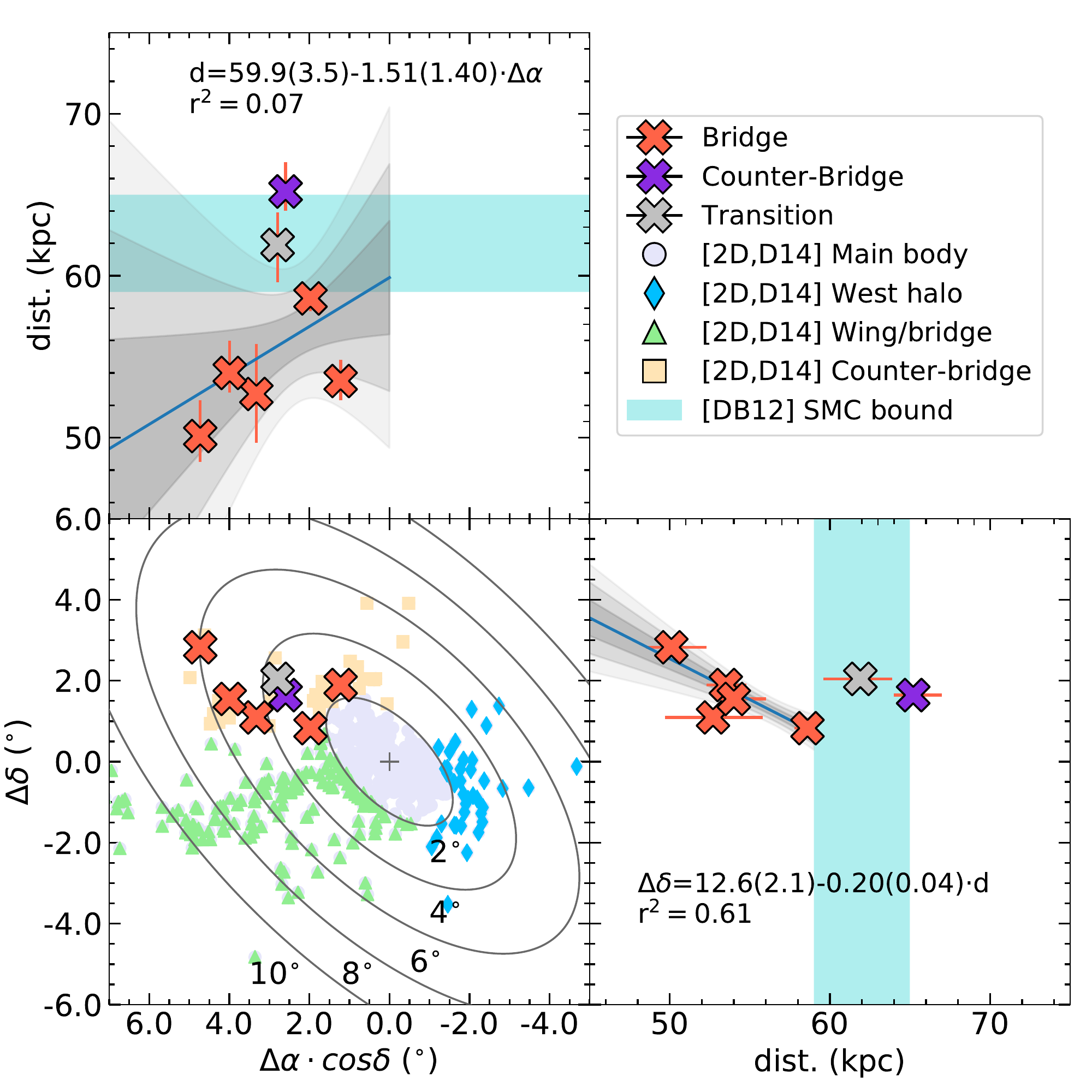}
    \caption{Position of the seven star clusters in relative equatorial coordinates as defined by D14, as seen in the planes ($\Delta\alpha$,$\Delta\delta$), ($\Delta\alpha$,dist), and ($\Delta\delta$,dist). Background points are all star clusters from \cite{bica+20}. The seven clusters are located in the proposed 'Counter-Bridge' 2D projected region by D14, but we now show that this region actually contains Bridge and Counter-Bridge clusters in 6D phase-space. The turquoise shaded area is a first approximation for the region bound to the SMC. Linear fits were performed for Bridge clusters only, considering the uncertainties. See Appendix \ref{app:3D} for a 3D view of the clusters and a movie is also provided online.}
    \label{fig:3Dposition}
\end{figure}

The SMC boundary along the line-of-sight defined by the simulation results of \cite{diaz+12} is between ${\sim}59$kpc and ${\sim}65$kpc, which we adopted here as a first approximation to identify five Bridge clusters and one (IC\,1708) Counter-Bridge cluster (see Table \ref{tab:results} and Fig. \ref{fig:3Dposition}). 
The seventh cluster (Bruck\,168) has a line-of-sight distance consistent with being bound to the SMC;
therefore, we placed this cluster in a transition region.
It is worth noting that the spheroid component of the Counter-Bridge (a.k.a. stars) is much less pronounced and less dense than the respective disc component (a.k.a. gas, see \citealp{omkumar+20}); as a result, it is not a surprise to find only a single representation of the Counter-Bridge in our cluster sample.
Moreover, there seems to be a trend in the sense that north-eastern Bridge clusters are also closer to us (see Fig. \ref{fig:3Dposition}), which supports the scenario that the foreground stellar population shapes the start of the Magellanic Bridge towards the LMC.

\begin{figure}[!htb]
    \centering
    \includegraphics[width=0.8\columnwidth]{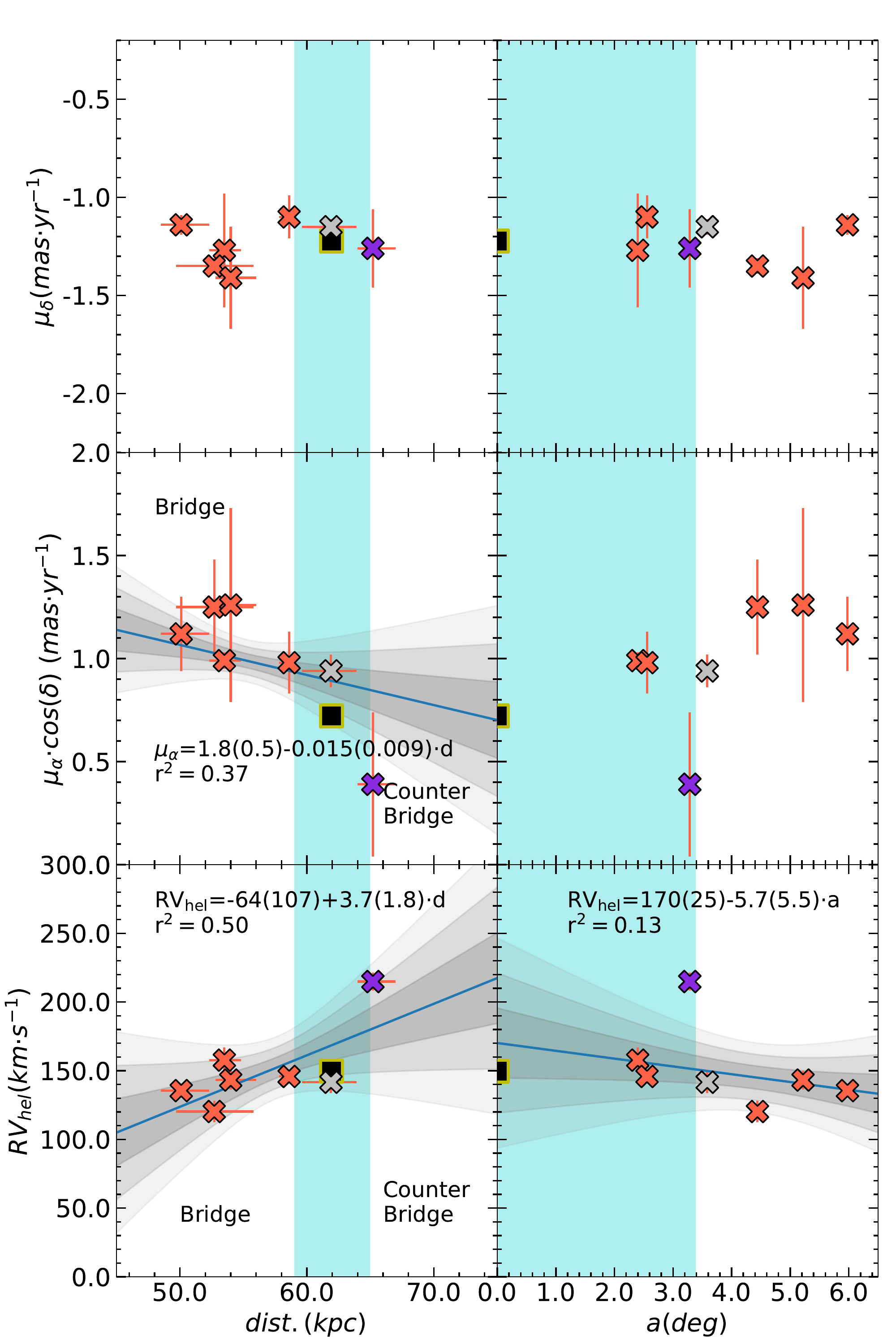}
    \caption{3D motion of the seven star clusters. 
    Point colours and shaded areas are the same as in Fig. \ref{fig:3Dposition}. The black square marks the SMC systemic motion and position (see Table \ref{tab:SMCLMCdata}). Linear fits were performed considering the uncertainties and using all clusters to allow direct comparison with other studies (see text). See Appendix \ref{app:3D} for a 3D view of the clusters and a movie is also provided online.
    }
    \label{fig:3Dmotion}
\end{figure}

Another crucial perspective is to analyse how these star clusters move with respect to the distance from the SMC in 3D space. Fig. \ref{fig:3Dmotion} shows how the RVs from the GMOS spectra and Gaia EDR3 proper motions vary with the line-of-sight distance and projected distance on sky $a$. Tidally stripped structures show a distance-velocity correlation, which characterises the Bridge and Counter-Bridge gas and stars very clearly (\citealp{gardiner+96}, Fig.11; \citealp{diaz+12}, Figs.9,15). In Fig. \ref{fig:3Dmotion}, we find a slope between RV$_{\rm hel}$ and the line-of-sight distance of $3.7\pm1.8{\rm km{\cdot}s^{-1}{\cdot}kpc^{-1}}$, which is in good agreement with the slope of $3.4{\rm km{\cdot}s^{-1}{\cdot}kpc^{-1}}$ found by \cite{gardiner+96} in their simulations and $4.0{\rm km{\cdot}s^{-1}{\cdot}kpc^{-1}}$ found by \cite{mathewson+88} based on Cepheids. This agreement is interesting because Cepheids trace the younger stellar populations, which would support a scenario where Bridge stars were formed in situ. However, our Bridge and Counter-Bridge sample is composed of clusters with ages ranging from ${\sim}1-4$Gyr, meaning that these clusters were formed before the LMC-SMC encounter ${\sim}150$Myr ago, which moved them from their original SMC orbits. Furthermore, \cite{hatzidimitriou+93} found a slope of $8{\rm km{\cdot}s^{-1}{\cdot}kpc^{-1}}$ based on red clump stars closer than 60kpc, which would trace intermediate-age and older populations. A larger sample of clusters is required to compare the correlation with this study. The correlation of RV$_{\rm hel}$ with the projected distance $a$ also reveals a trend in the sense that the Bridge clusters that are farther away from the SMC projected centre are also closer to us and dragging behind towards the LMC. These results are consistent again with one branch of the Magellanic Bridge starting in the north-eastern SMC. The Counter-Bridge also starts in the same region, but so far we only have one point, and a larger sample would help constrain this tidal tail. Gaia EDR3 proper motions show relatively large error bars (Fig.\ref{fig:3Dmotion}); nevertheless, the bulk of Bridge clusters are indeed separated from the Counter-Bridge cluster in $\mu_{\alpha}$, whereas $\mu_{\delta}$ is roughly constant.
This is consistent with the scrutiny of \cite{omkumar+20} on the simulations by \cite{diaz+12} and also consistent with the Gaia EDR3 proper motion distribution of the Magellanic System (see Appendix \ref{app:vpd}).

Last but not least, concerning the eighth dimension analysed here, we report that the metallicity of the five Bridge clusters ($\langle{\rm[Fe/H]_{CMD}^{Bridge}}\rangle= -0.69\pm0.12$ dex; $\langle{\rm[Fe/H]_{CaT}^{Bridge}}\rangle= -0.90\pm0.06$ dex) is ${\sim}0.2-0.3$ dex, that is, $\sim 2\sigma$ higher than the Counter-Bridge cluster (${\rm[Fe/H]_{CMD}^{C.-Bridge}} = -1.02^{+0.05}_{-0.10}$ dex, ${\rm[Fe/H]_{CaT}^{C.-Bridge}} = -1.11\pm0.06\pm0.17$ dex). Moreover, the Bridge clusters are all older than the Counter-Bridge cluster.
 This suggests differences in metallicities per SMC region as indicated by \cite{crowl+01} and endorses further investigation.
A possible metallicity gradient would not eliminate this difference because more distant Counter-Bridge clusters would have similar metallicities \citep{parisi+15} or be even more metal-poor (D14).

\section{Conclusions}

The gaseous complex structure of the Magellanic System and its stellar counterpart, in particular around the SMC, has been observed and simulated by many previous works. We
present an analysis of this question using star clusters observed within the
VISCACHA survey. The advantage of this approach is that we are able to
describe the Magellanic Bridge and Counter-Bridge in the 6D phase-space
plus age and metallicity. We have reached ${\sim}1-6$\% precision in distance, ${\sim}0.5-8$\% precision in the mean RV, not to mention ${\sim}4-20$\% precision in age, and ${\sim}0.03-0.22$ dex and  ${\sim}0.04-0.12$ dex precision in the mean photometric and spectroscopic metallicities. 

From the seven clusters analysed here, we found that five belong to the Magellanic Bridge, while one belongs to the Counter-Bridge and another is located at an intermediate region between the two tidal tails. Six-dimensional phase-space vectors of these clusters are consistent with the predictions from the simulations by \cite{diaz+12}, which confirms their unbound current situation. These clusters are 1-4Gyr old, therefore, they were formed before the SMC-LMC close encounter that generated the SMC tidal tails and moved these clusters away from the SMC main body.

The 2D projected SMC Counter-Bridge region as defined by D14,D16 contains a mix of Bridge and Counter-Bridge clusters in a five-to-one ratio. Consequently, the Magellanic Bridge also has a branch starting from the north-eastern SMC, in addition to the eastern SMC wing and south-eastern RR Lyrae bridge. 

The present sample gives important hints on a likely scenario for the formation of
a structure in the clouds, in particular the Bridge and the Counter-Bridge.
A larger sample is expected to further constrain the perturbed outskirts of the SMC.

\begin{acknowledgements}
B.D. and M.C.Parisi acknowledge S.Vasquez for providing his code to measure CaT equivalent widths.\\
Based on observations obtained at the Southern Astrophysical Research (SOAR) telescope, which is a joint project of the Minist\'erio da Ci\^encia, Tecnologia, e Inova\c c\~ao (MCTI) da Rep\'ublica Federativa do Brasil, the U.S. National Optical Astronomy Observatory (NOAO), the University of North Carolina at Chapel Hill (UNC), and Michigan State University (MSU).\\
Based on observations obtained at the international Gemini Observatory, a programme of NSF’s NOIRLab, which is managed by the Association of Universities for Research in Astronomy (AURA) under a cooperative agreement with the National Science Foundation. on behalf of the Gemini Observatory partnership: the National Science Foundation (United States), National Research Council (Canada), Agencia Nacional de Investigaci\'{o}n y Desarrollo (Chile), Ministerio de Ciencia, Tecnolog\'{i}a e Innovaci\'{o}n (Argentina), Minist\'{e}rio da Ci\^{e}ncia, Tecnologia, Inova\c{c}\~{o}es e Comunica\c{c}\~{o}es (Brazil), and Korea Astronomy and Space Science Institute (Republic of Korea). Programme ID: GS-2018B-Q-208, GS-2018B-Q-302.\\
This work has made use of data from the European Space Agency (ESA) mission
{\it Gaia} (\url{https://www.cosmos.esa.int/gaia}), processed by the {\it Gaia}
Data Processing and Analysis Consortium (DPAC,
\url{https://www.cosmos.esa.int/web/gaia/dpac/consortium}). Funding for the DPAC
has been provided by national institutions, in particular the institutions
participating in the {\it Gaia} Multilateral Agreement.\\
This research was partially supported by the Argentinian institutions CONICET, SECYT (Universidad Nacional de Córdoba), Universidad Nacional de La Plata and Agencia Nacional de Promoción Científica y Tecnológica (ANPCyT). \\
This study was financed in
part by the Coordenação de Aperfeiçoamento de Pessoal de
Nível Superior - Brasil (CAPES) - Finance Code 001.\\
A.P.V. and S.O.S. acknowledge the DGAPA-PAPIIT grant IG100319.\\
D.G. gratefully acknowledges support from the Chilean Centro de Excelencia en Astrof\'{i}sica
y Tecnolog\'{i}as Afines (CATA) BASAL grant AFB-170002.
D.G. also acknowledges financial support from the Direcci\'on de Investigaci\'on y Desarrollo de
la Universidad de La Serena through the Programa de Incentivo a la Investigaci\'on de
Acad\'emicos (PIA-DIDULS).\\
R.A.P.O. and S.O.S. acknowledge the FAPESP PhD fellowships nos. 2018/22181-0 and 2018/22044-3\\
 The authors thank the referee for the comments that improved this letter.
\end{acknowledgements}

   \bibliographystyle{aa} 
   \bibliography{bibliography} 

\appendix

\section{Supporting material}
\subsection{Isochrone fitting and spectroscopic membership}
\label{app:CMDspec}

Here, we present the CMD isochrone fitting and spectroscopic membership selection. 

\begin{figure*}
    \centering
    \includegraphics[width=0.3\textwidth]{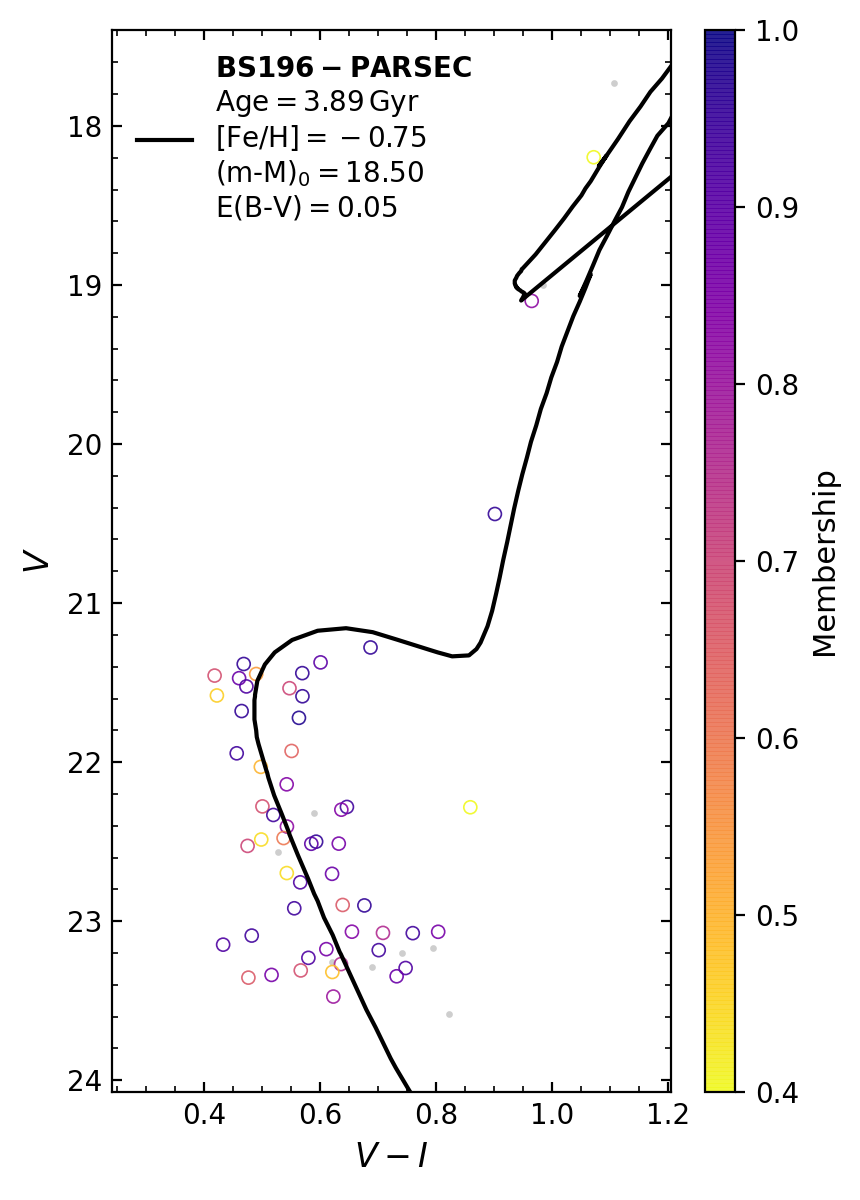}
    \includegraphics[width=0.3\textwidth]{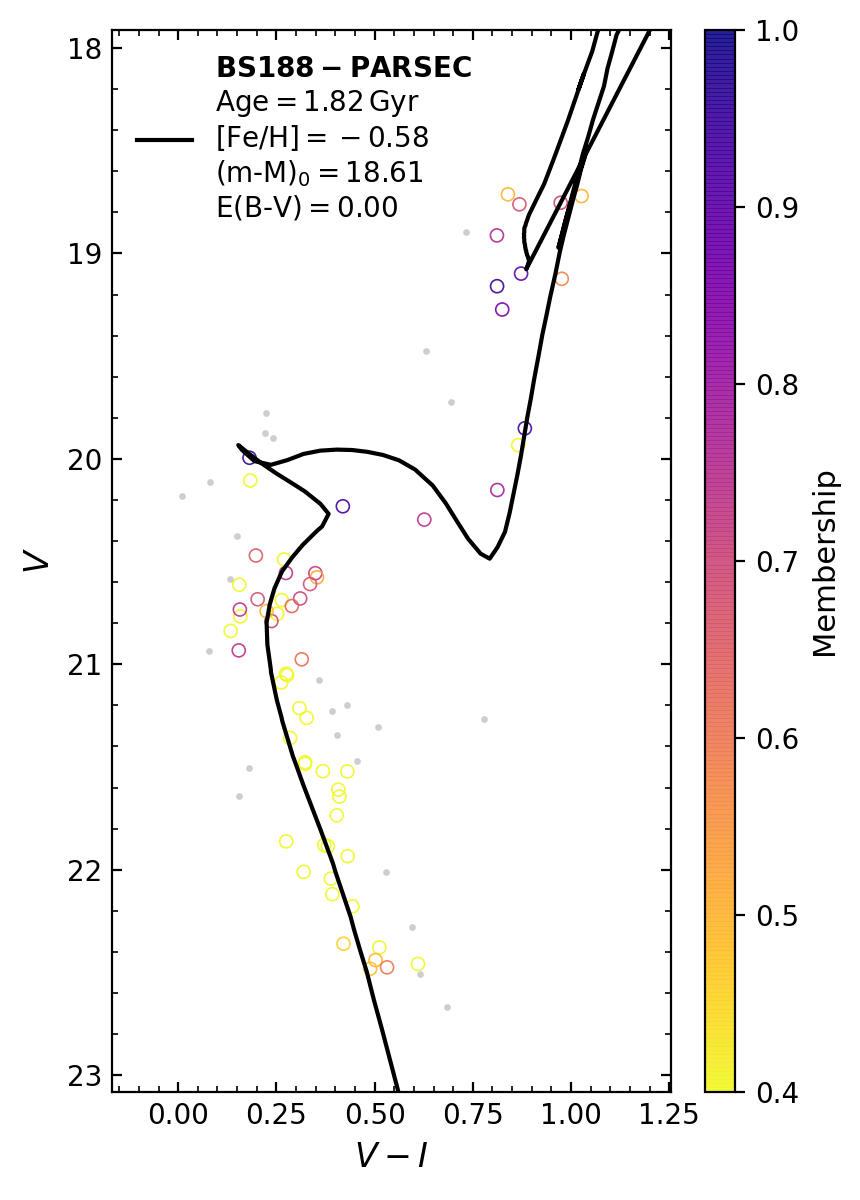}
    \includegraphics[width=0.3\textwidth]{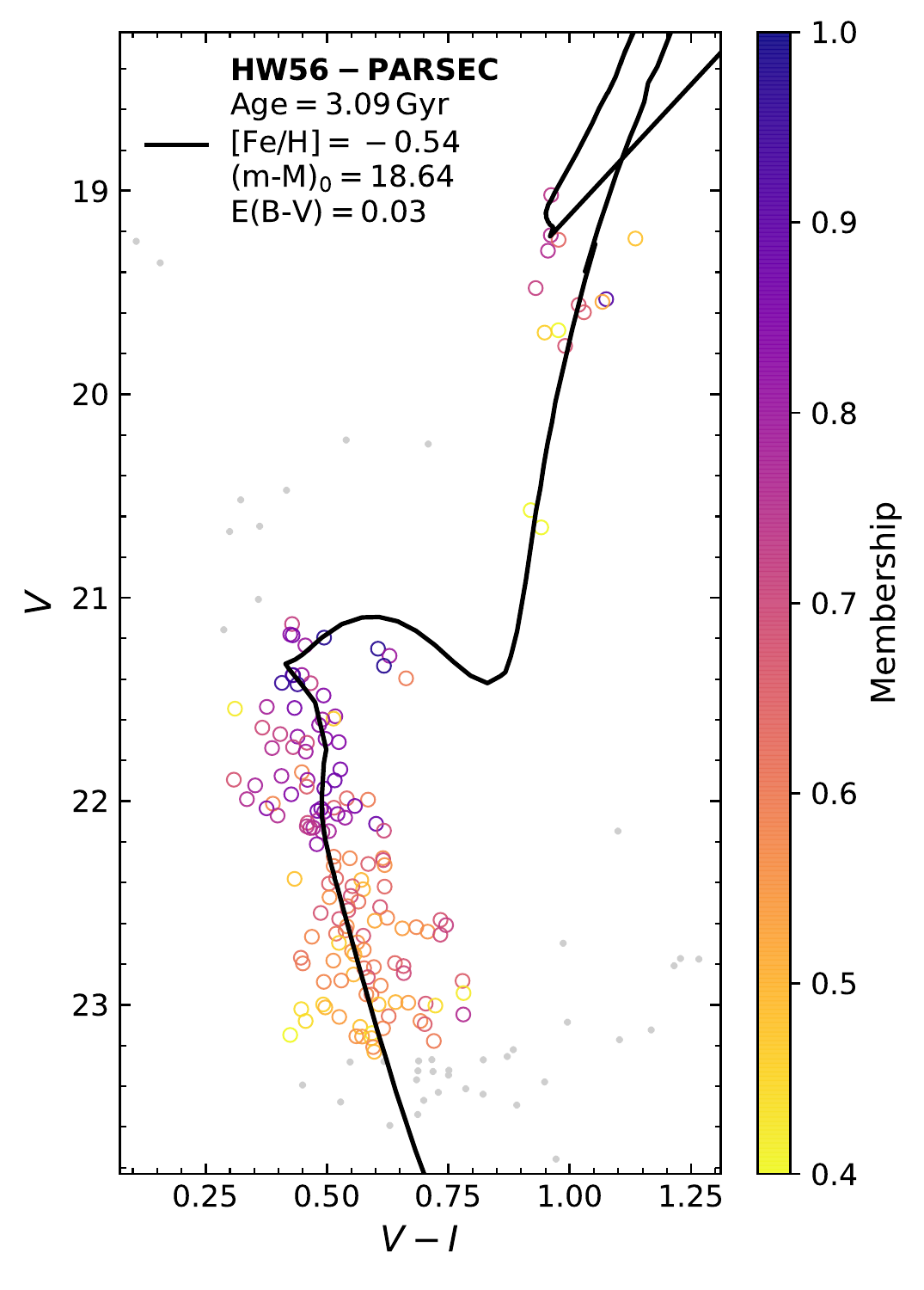}
    \includegraphics[width=0.3\textwidth]{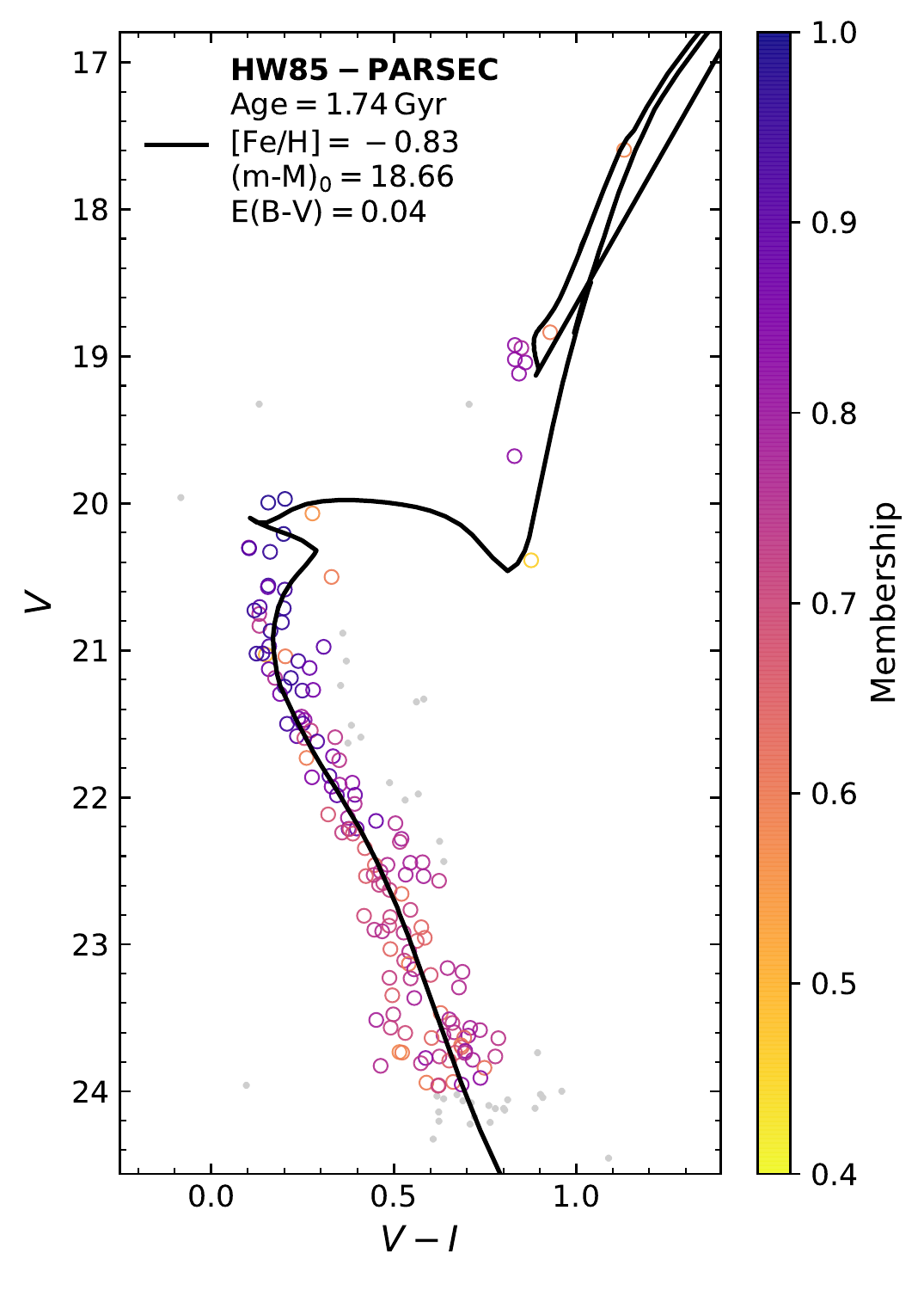}
    \includegraphics[width=0.3\textwidth]{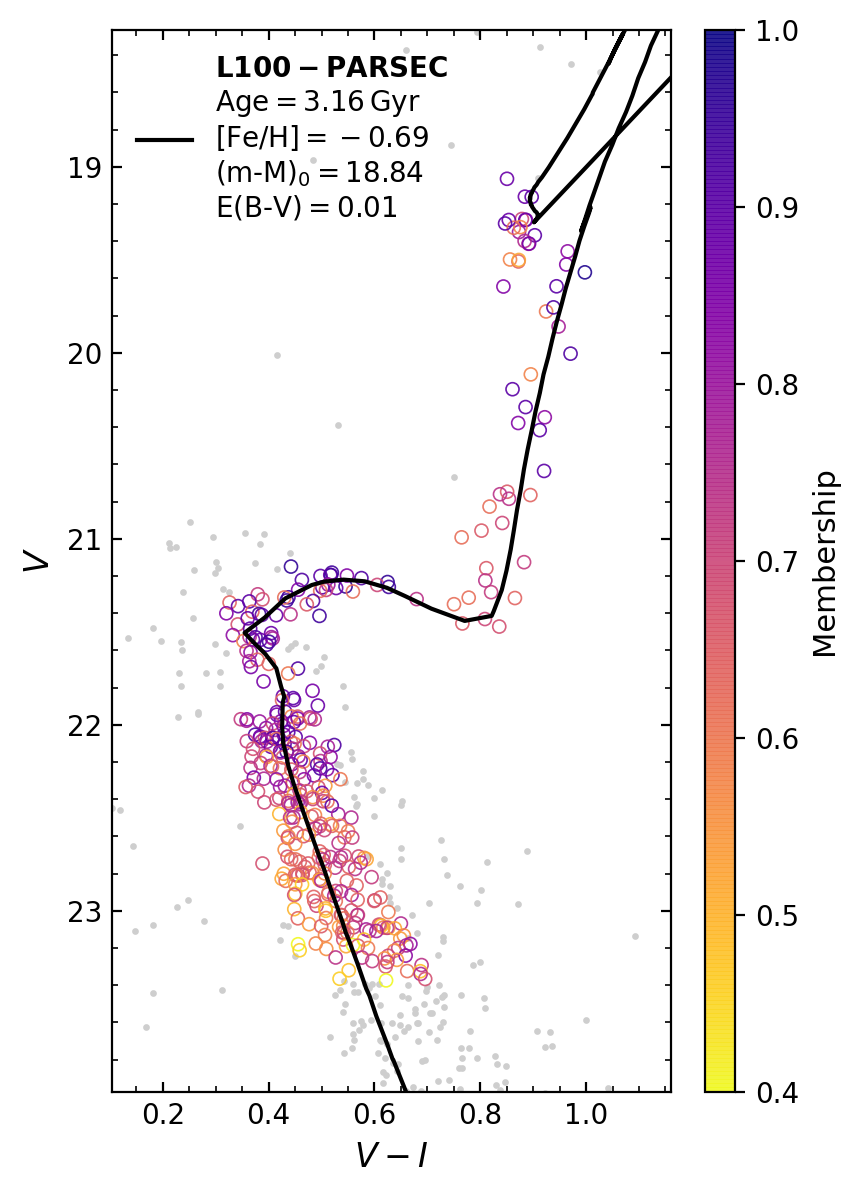}
    \includegraphics[width=0.3\textwidth]{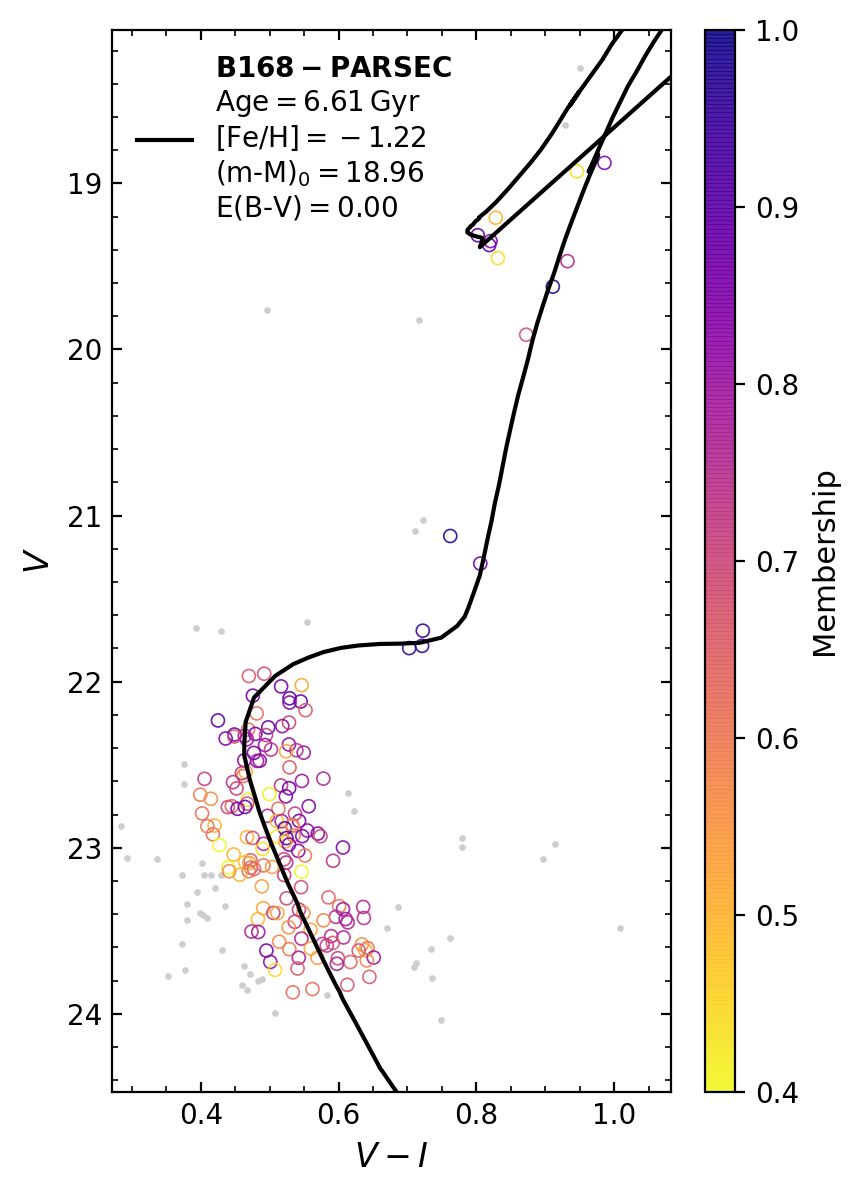}
    \includegraphics[width=0.3\textwidth]{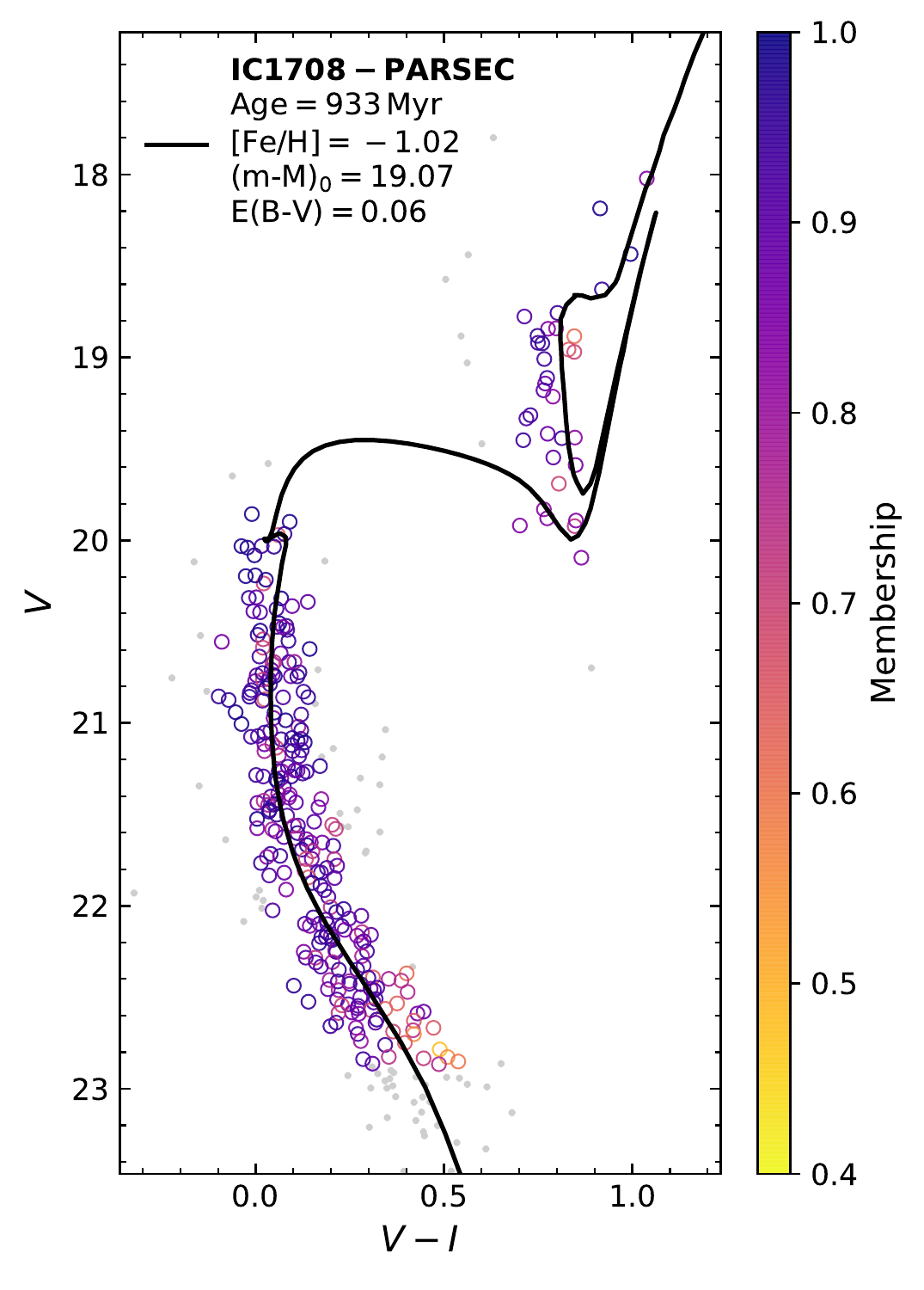}
    \caption{Statistically decontaminated CMD with the best isochrone fitting for each cluster analysed here, using the SIRIUS code. Grey dots represent field stars, whereas the circles are probable cluster member stars, colour-coded by their membership probability.}
    \label{fig:CMDfit_appendix}
\end{figure*}

\begin{figure*}[!h]
    \centering
    $\begin{array}{cp{1cm}c}
    \includegraphics[width=0.6\columnwidth]{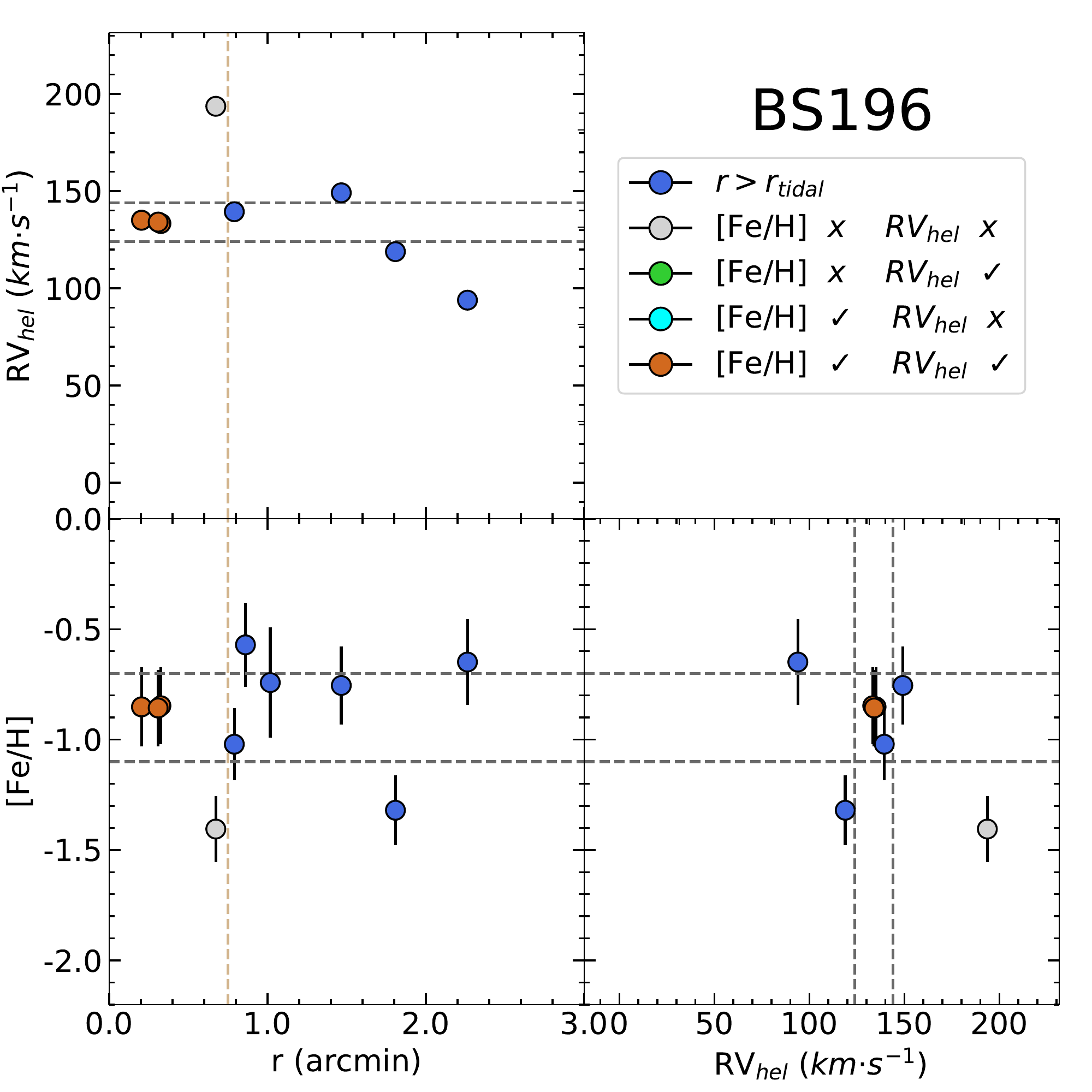} & & 
    \includegraphics[width=0.6\columnwidth]{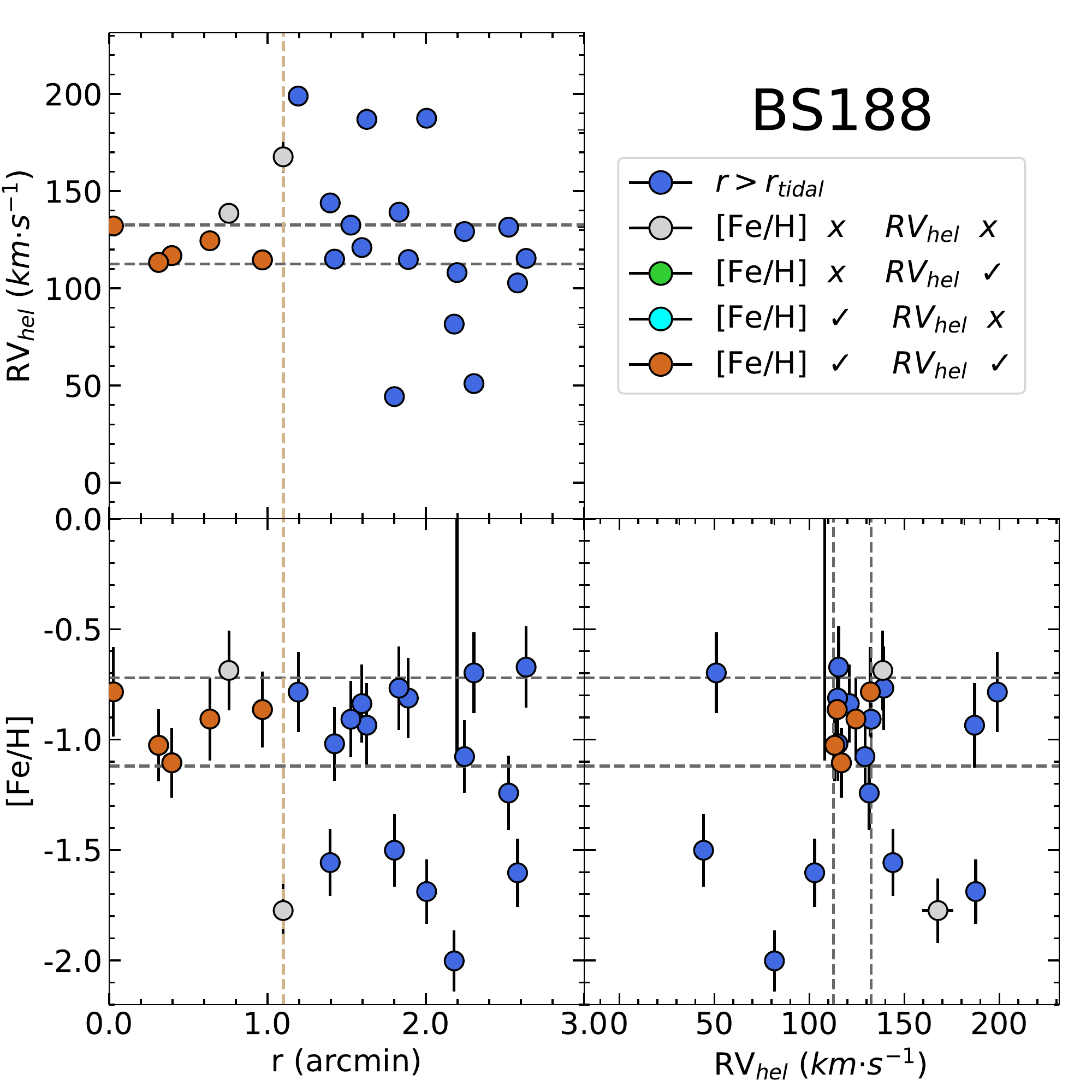}
    \end{array}$
    $\begin{array}{cp{1cm}c}
    \includegraphics[width=0.6\columnwidth]{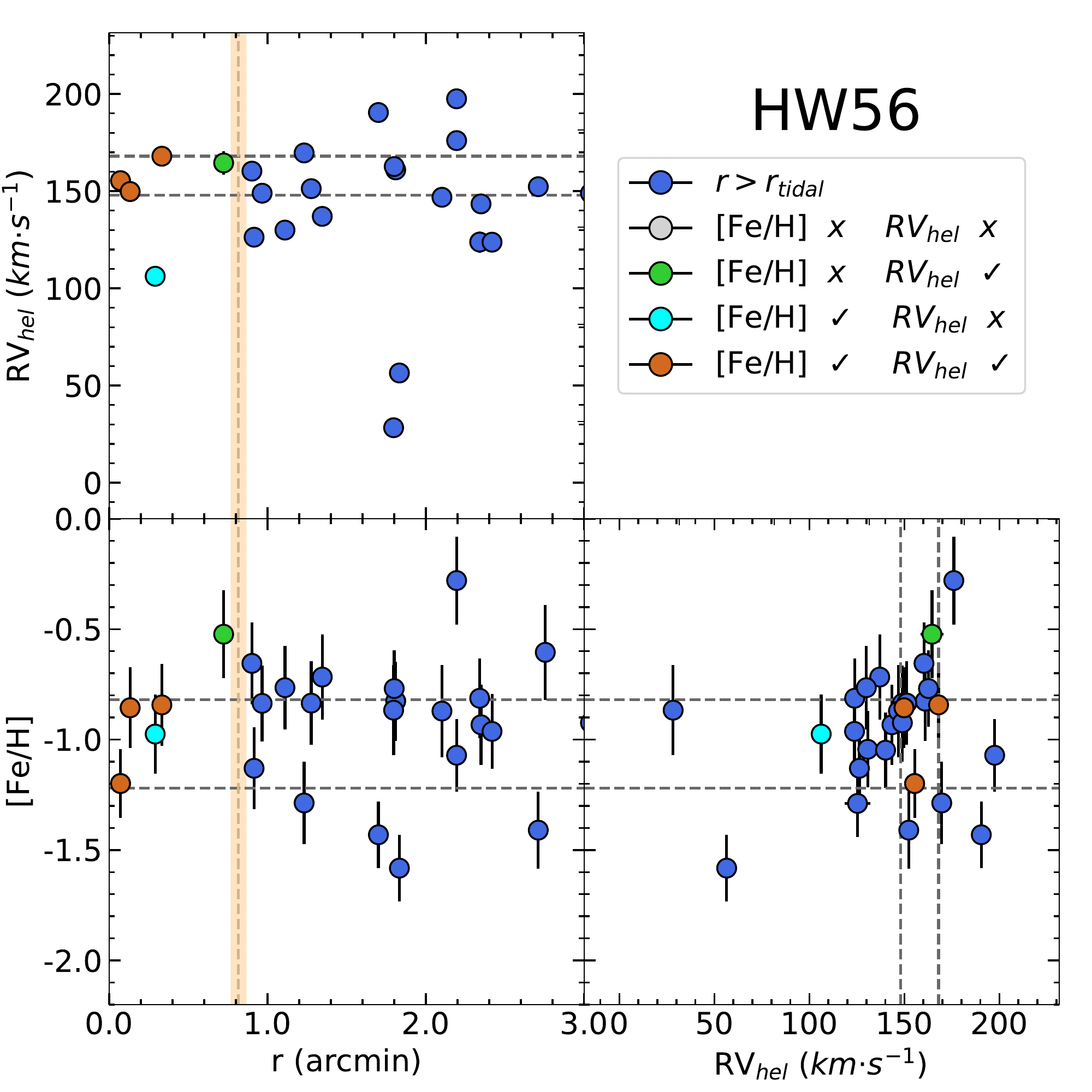} & & 
    \includegraphics[width=0.6\columnwidth]{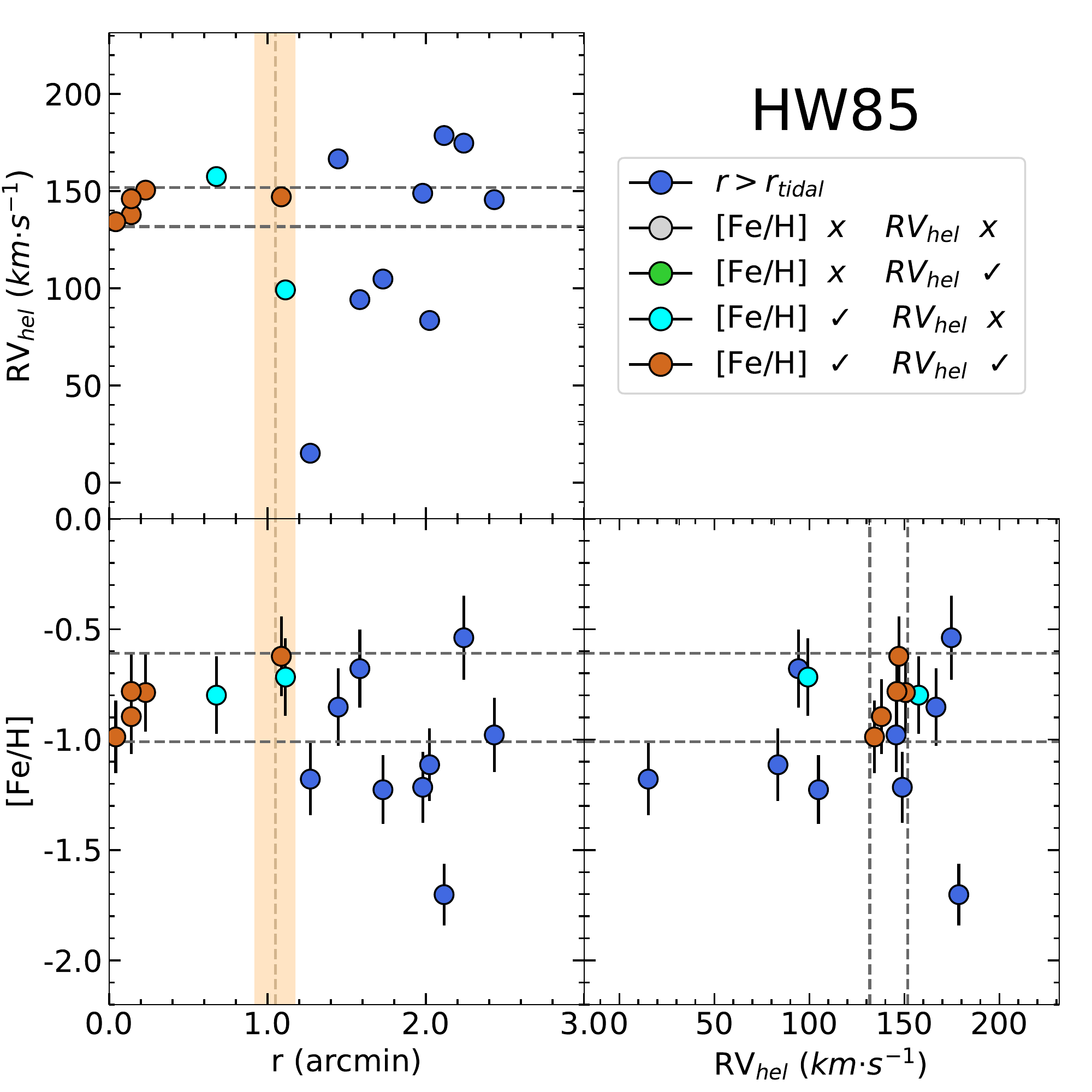}
    \end{array}$
    $\begin{array}{cp{1cm}c}
    \includegraphics[width=0.6\columnwidth]{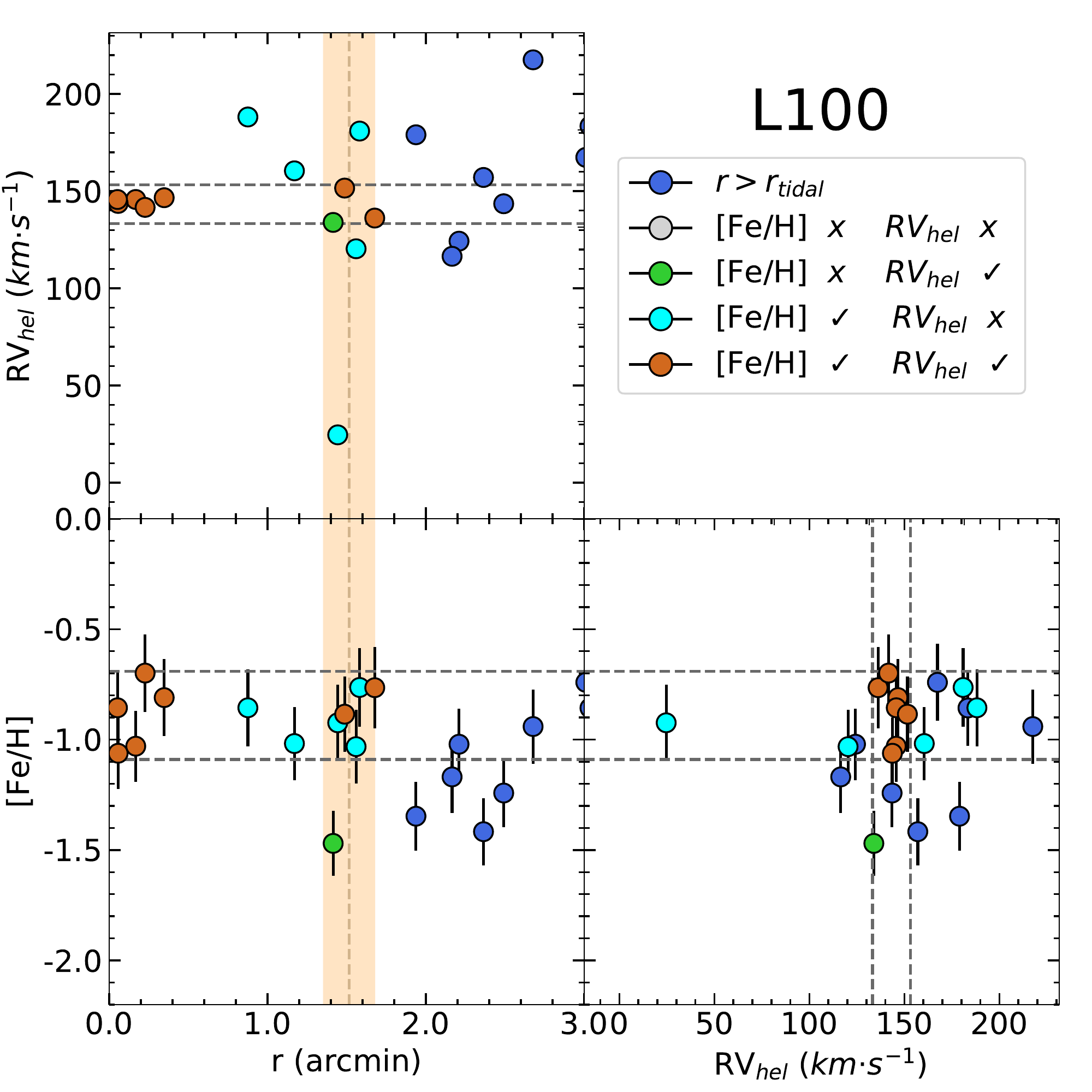} & & 
    \includegraphics[width=0.6\columnwidth]{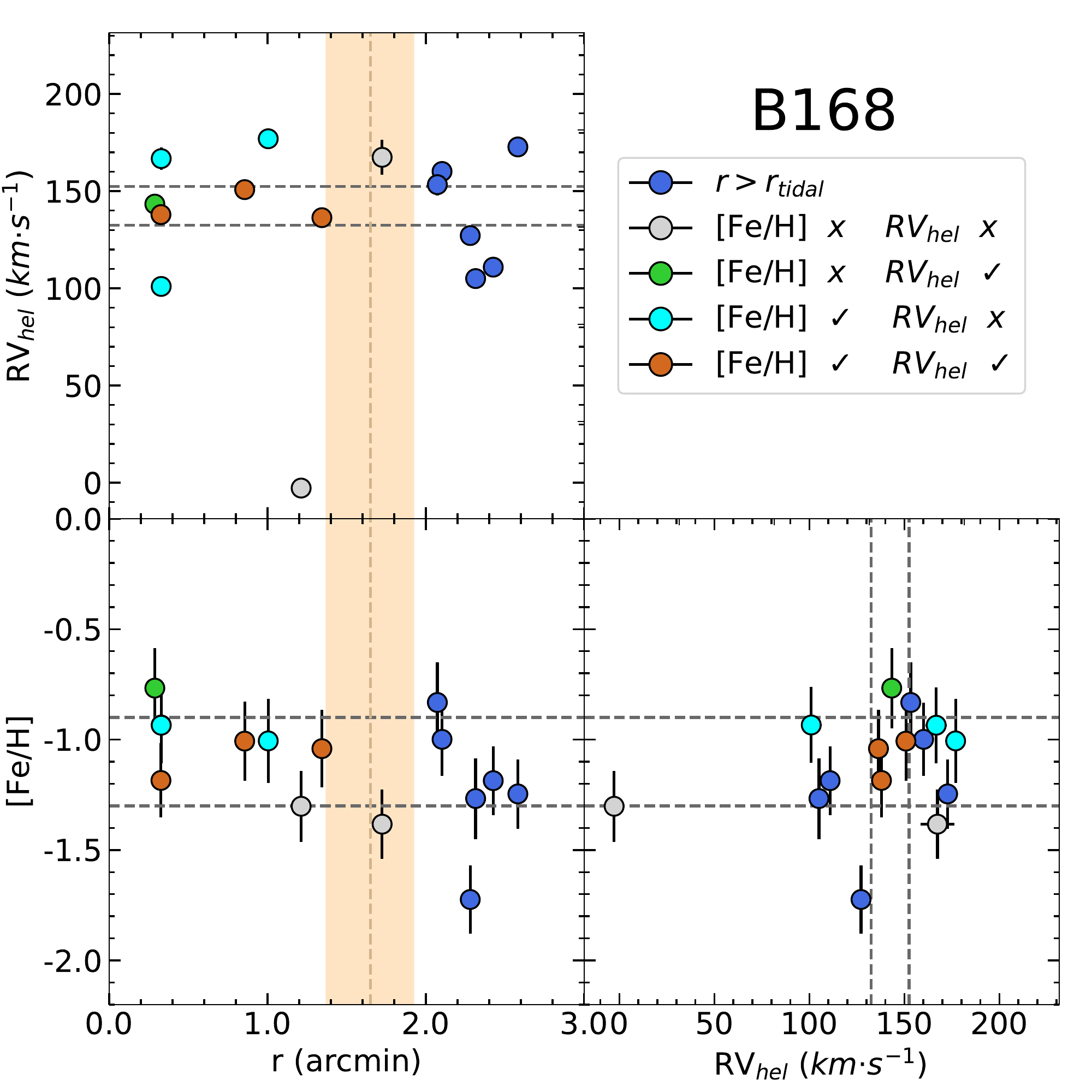}
    \end{array}$
    $\begin{array}{c}
    \includegraphics[width=0.6\columnwidth]{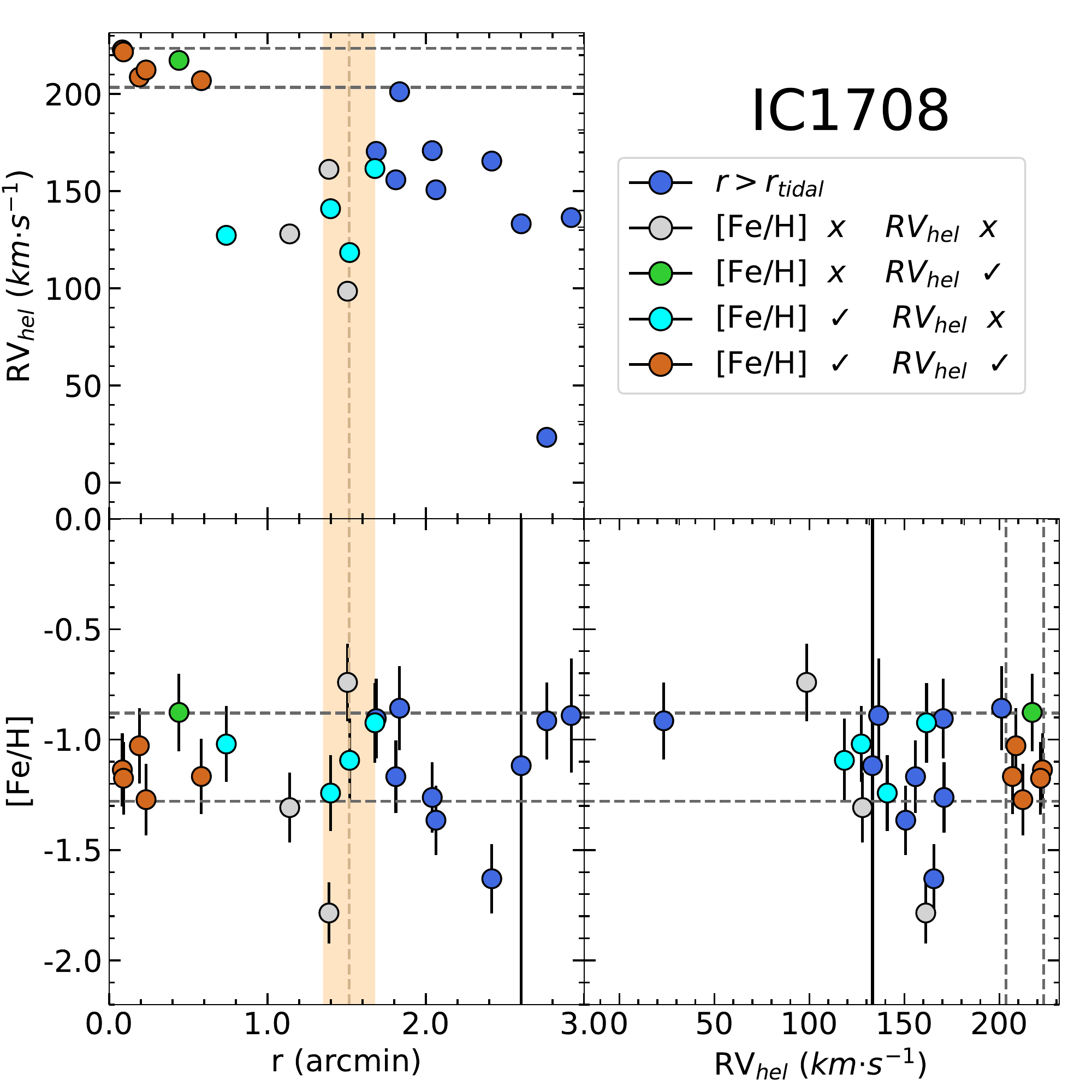}
    \end{array}$
    \caption{Membership selection of cluster stars with spectroscopic information. The shaded area marks the cluster tidal radius $\pm1\sigma$ from \cite{santos+20}, otherwise only a line represents twice the visual radius by \cite{bica+20}. The limits in [Fe/H] and $RV_{hel}$ are 0.2 dex and 10km$\cdot$s$^{-1}$ around the group of innermost stars.}
    \label{fig:RVsel_appendix}
\end{figure*}

\subsection{Vector-point diagrams from Gaia EDR3}
\label{app:vpd}

\begin{figure*}[!htb]
    \centering
    \includegraphics[width=\textwidth]{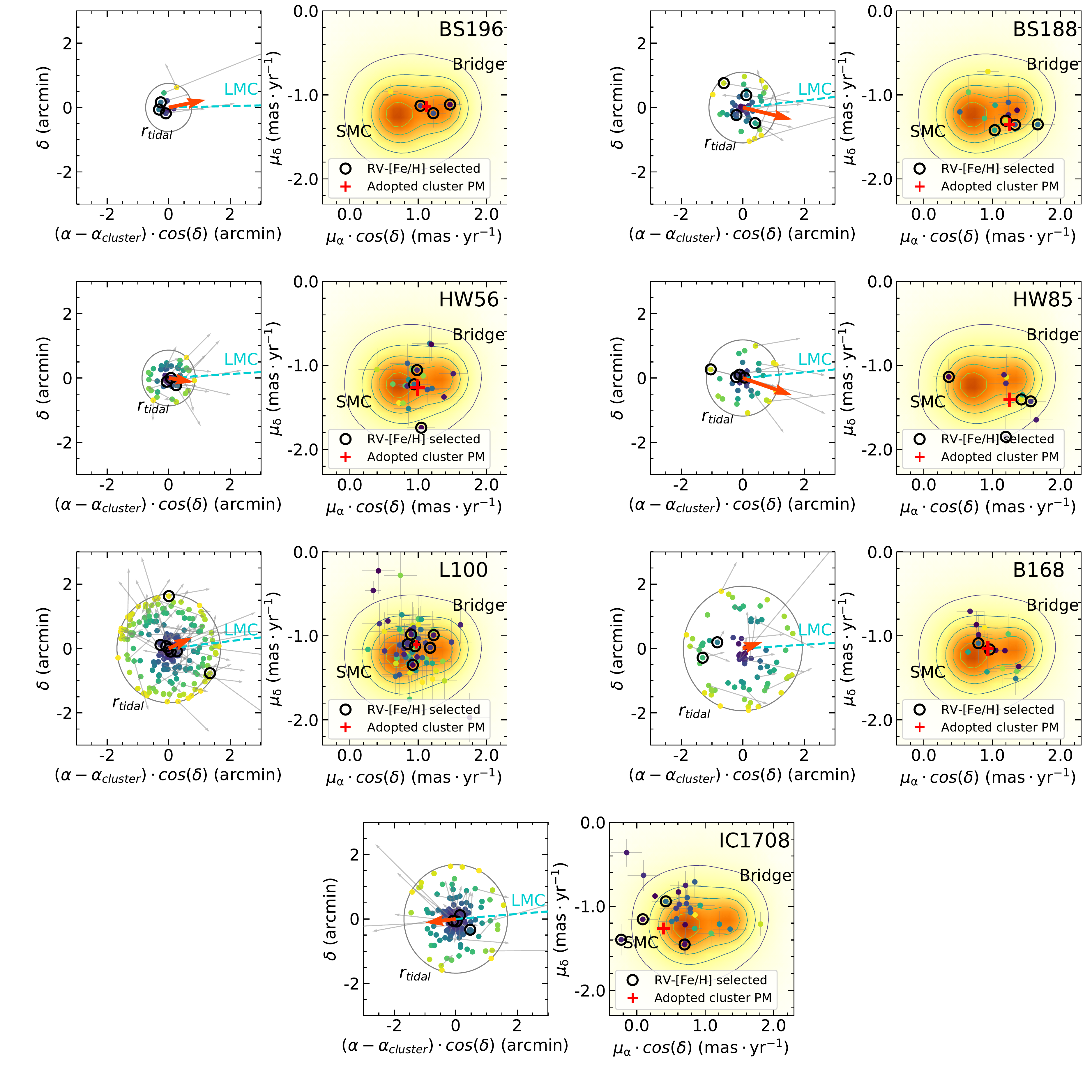}
    \caption{Vector-point diagram from Gaia EDR3 proper motions for the seven clusters. {\it Left panel:} On sky distribution of all Gaia stars within the tidal radius of the cluster. Colours indicate the distance from the cluster centre. Grey arrows are Gaia EDR3 good-quality proper motions subtracted from the SMC mean proper motion. The red thick arrow is the average of the selected member stars highlighted with black circles. The direction of the LMC is indicated by the turquoise line. {\it Right panel:} The background density plot represents the locus of the SMC and Bridge based on a sample of stars (see text for details). The points and their error bars are the equivalent from the left panel.}
    \label{fig:VPD}
\end{figure*}

We present the vector-point diagram (VPD) showing the position of the SMC, Bridge, 
and all selected cluster stars in Fig. \ref{fig:VPD}. In all cases, an initial quality filter was applied \citep[see][]{vasiliev+18} selecting only stars with $\sigma_{\mu_{\alpha}} < 0.2\ {\rm mas\cdot yr^{-1}}$, $\sigma_{\mu_{\delta}} < 0.2\ {\rm mas\cdot yr^{-1}}$, and $\pi < 3\cdot\sigma_{\pi}$, that is to say a parallax consistent with zero. These criteria resulted in a single star for four clusters; therefore, we relaxed the constraint on proper motions errors from 0.2 to 0.3 ${\rm mas\cdot yr^{-1}}$ for a better compromise between statistics and uncertainties. Only one star of B\,168 lies outside the plot area and it was considered an outlier.
The locus of the SMC 
is represented by a sample of stars located within $0.5^{\circ}$ from
its optical 
centre. The locus of the Bridge is represented by all good-quality stars within $0.5^{\circ}$ around the cluster BS225 position, which is a random Bridge cluster far away from the SMC centre \citep{bica+15}. The relative density of stars between the SMC and
Bridge was optimised for best visualisation of their positions in the VPD only. The membership selection of stars for each cluster does not use proper motion information, only RVs and metallicities as described above. The stars indicated in Fig. \ref{fig:VPD} are those good-quality stars from Gaia that match the selected member stars for each cluster. The proper motions of each cluster is the weighted average of the selected Gaia stars. The systematic uncertainty of $\sigma_{\mu} = 0.01 {\rm mas\cdot yr^{-1}}$ given by \cite{gaiaMC20} is negligible in comparison with the uncertainties reported in Table \ref{tab:results}. Looking at Fig.\ref{fig:VPD}, we confirm that the Bridge clusters are pointing towards the LMC, which is consistent with the Bridge,
whereas the Counter-Bridge cluster is moving in the opposite direction.

\subsection{3D view of the SMC clusters}
\label{app:3D}

\begin{figure}[!htb]
    \centering
    \includegraphics[width=\columnwidth]{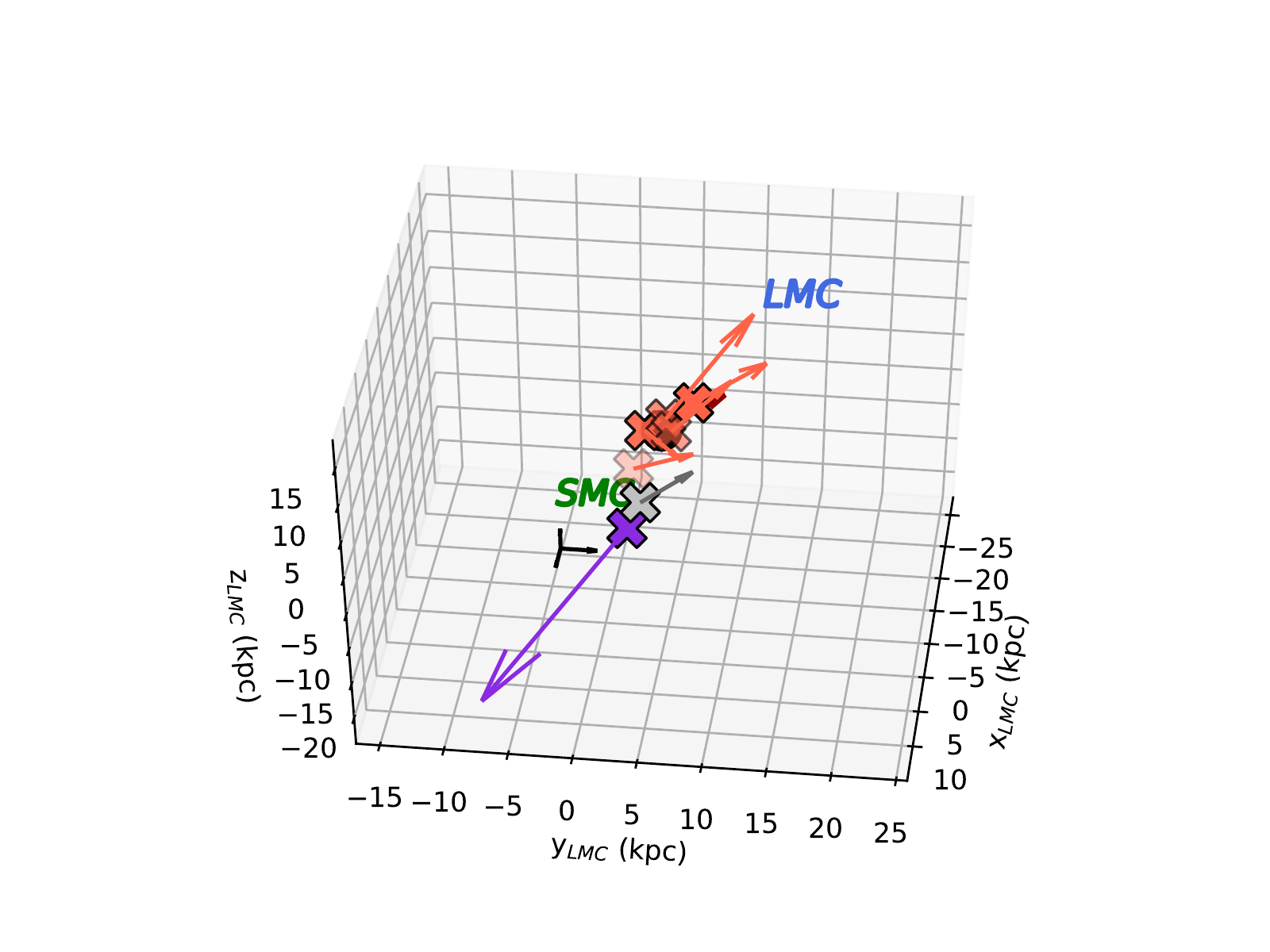}
    \caption{Phase-space vectors of the seven clusters analysed in this work in a 3D Cartesian system centred at the SMC as described in the text. 
    Arrows are the velocities relative to the SMC mean velocity. 
    Colours and symbols are the same as in Fig. \ref{fig:3Dposition}. The average velocity of the five Bridge clusters is shown in dark red. The black arrows are a reference scale representing 10 ${\rm km}\cdot {\rm s}^{-1}$.  A movie is available as online material at https://www.aanda.org/.}
    \label{fig:3Dcart}
\end{figure}

In order to provide an additional visualisation of the results, we calculated the Cartesian coordinates of the clusters in a reference system, with an origin at the LMC, the z-axis pointing towards us, the y-axis towards the north, and the x-axis towards the  west. We applied the Eqs. 1,2,3,5 from \cite{vdM+01}. The velocity vectors in this Cartesian system were calculated using Eqs. 3,6,7,8 from \cite{vdM+02}. We also calculated the velocity vectors for the mean motion of the 
SMC and subtracted its mean position and velocity
from all clusters and the LMC to finally produce Fig. \ref{fig:3Dcart} with positions and velocities relative to the SMC. The adopted mean position and velocities of the SMC and LMC are given in Table \ref{tab:SMCLMCdata}. It is very clear that cluster IC\,1708 and the Bridge clusters are moving in opposite directions, roughly aligned with the SMC-LMC direction, as it can also be seen in Fig.\ref{fig:VPD}.

\begin{table}[!htb]
    \centering
    \footnotesize
    \caption{Mean parameters adopted for the SMC and LMC.}
    \label{tab:SMCLMCdata}
    \begin{tabular}{p{1.4cm}p{1.3cm}|rrc}
    \hline
    \noalign{\smallskip}
  \multicolumn{1}{c}{param.} &
  \multicolumn{1}{c|}{unit} &
  \multicolumn{1}{c}{SMC} &
  \multicolumn{1}{c}{LMC} &
  \multicolumn{1}{c}{ref.} \\
    \noalign{\smallskip}
    \hline\hline
    \noalign{\smallskip}
    $\alpha(J2000)$  & hh:mm:ss  & 00:53:45  & 05:19:31  & 1,2     \\
    $\delta(J2000)$  & dd:mm:ss  & $-$72:49:43 & $-$69:35:34               & 1,2     \\
    (m-M)      & mag        & $18.96\pm0.02$                    & $18.49\pm0.09$                     & 3,4    \\
    RV$_{\rm helio}$ & ${\rm km\cdot s^{-1}}$  & $149.6\pm0.8$ & $262.2\pm3.4$  &  5,6     \\
    $\mu_{\alpha}\cdot{ }cos(\delta)$ & ${\rm mas{\cdot}yr^{-1}}$ &  $0.721\pm0.024$        & $1.910\pm0.020$                               &  5,7     \\
    $\mu_{\delta}$  & ${\rm mas{\cdot}yr^{-1}}$    &   $-1.222\pm0.018$                  & $0.229\pm0.047$                               &  5,7     \\
    \noalign{\smallskip}
    \hline
    \end{tabular}
    \tablefoot{1. \cite{crowl+01}; 2. \cite{vdM+14}; 3. \cite{degrijs+15}; 4. \cite{degrijs+14}; 5. \cite{deleo+20}; 6. \cite{vdM+02}; 7. \cite{kallivayalil+13}.}
\end{table}

\end{document}